\documentclass[journal,10pt]{IEEEtran}

\usepackage{amsmath}
\usepackage{bbold}
\usepackage{enumerate}
\usepackage{cite}
\usepackage{xcolor}
\usepackage{amssymb}
\usepackage{graphicx}
\usepackage{algorithm}
\usepackage{algpseudocode}
\usepackage{bbm}
\usepackage{makecell}
\usepackage{graphics}
\usepackage{subfigure}
\usepackage{epsfig}
\usepackage{epstopdf}
\usepackage{arydshln}
\newtheorem{theorem}{Theorem}{}
\newtheorem{proposition}{Proposition}{}
\newtheorem{lemma}{Lemma}{}
\newtheorem{assumption}{Assumption}
\newtheorem{corollary}{Corollary}{}
\newtheorem{remark}{Remark}{}
\newtheorem{definition}{Definition}{}

\begin{document}

\title{Cooperative Control of Multi-Channel Linear Systems with Self-Organizing Private Agents}

\author{Peihu~Duan, Tao~Liu, Yuezu~Lv, and Guanghui Wen 
\thanks{P. Duan and T. Liu are with the Department of Electrical and Electronic Engineering, The University of Hong Kong, Pokfulam, Hong Kong SAR, China.  E-mails: phduan@hku.hk (P. Duan), taoliu@eee.hku.hk (T. Liu). \emph{(Corresponding author: Tao Liu)}  }

 \thanks{Y. Lv is with Advanced Research Institute of Multidisciplinary Sciences, Beijing
  Institute of Technology, Beijing 100081, China. E-mail: yzlv@bit.edu.cn. }

\thanks{G. Wen is with the Department of Systems Science, School of Mathematics, Southeast University, Nanjing 211189, China. E-mail: wenguanghui@gmail.com.}
}

\maketitle

\begin{abstract}
  Cooperative behavior design for multi-agent systems with collective tasks is a critical issue in promoting swarm intelligence. This paper investigates cooperative control for a multi-channel system, where each channel is managed by an agent expected to self-organize a controller to stabilize the system collaboratively by communicating with neighbors in a network. Integrating a state decomposition technique and a fusion approach, a fully distributed privacy-preserving mechanism is proposed to shield agents' private information from neighbors' eavesdropping. Moreover, the cost of introducing the privacy-preserving mechanism and the benefit of adding more channels to the system are quantitatively analyzed. Finally, comparative simulation examples are provided to demonstrate the effectiveness of the theoretical results.
\end{abstract}

\begin{IEEEkeywords}
Cooperative control, Privacy preservation, Self-organization, Multi-agent system
\end{IEEEkeywords}

\section{Introduction} \label{sec1}
Over the past two decades, multi-agent systems have experienced tremendous growth in both theoretical research and practical deployment, ranging from transportation systems, industrial equipments to large-scale infrastructure \cite{pipattanasomporn2009multi,cao2012overview,marino2017distributed}. Compared with a single system, multi-agent systems possess great superiority in rich functionality and system robustness. One critical issue for multi-agent systems is highly efficient teamwork/cooperation in a distributed manner, where every individual agent performs a subtask \cite{olfati2007consensus}. For this issue, developing an effective cooperative control strategy is a requisite. In addition, accompanied by local interaction, sensitive private information of each agent is more likely to be recovered by neighbors, which jeopardizes the security of the whole system and hinders the collaboration of other agents \cite{roman2013features}. Therefore, privacy and security of multi-agent systems also deserve sufficient attention.

\subsection{Related Works} \label{sec1.1}
Motivated by the well-developed system theory and graph theory, many cooperative control strategies have been proposed for multi-agent systems, such as consensus  \cite{ren2005consensus}, containment  \cite{meng2010distributed}, formation  \cite{oh2015survey}, and synchronization \cite{liu2010exponential}. These strategies mainly establish on a hypothesis that all agents have a common or similar control objective, e.g., converging to a consistent view of states. However, in many practical applications, such as collaborative manipulation/transportation in smart factories/warehouses, different agents may have different and complementary capabilities such that they need to execute nonidentical subtasks decomposed from a collective task. For these cases, the above strategies are not applicable. As a remedy, the plant manipulated/transported by the multi-agent system can be regarded as a dynamical multi-channel system, where each channel is managed by an agent \cite{lavaei2009overlapping}. Specifically, each agent measures partial system state information and exerts a partial system input through the corresponding channel. Subsequently, the key matter becomes how to design the control input for each channel, i.e., the required input exerted on the plant by each agent. This paper focuses on this cooperative control problem.


To address the above problem of interest, existing strategies can be divided into centralized, decentralized, and distributed ones. Centralized strategies utilize a central authority to collect all system information to design  control inputs for all channels \cite{bertsekas1995dynamic,zadeh2008linear,li2017cooperative}. In this sense, the original cooperative problem can be converted into a standard stabilization problem for a single dynamical system. Although centralized strategies are able to achieve the optimal system performance for any pre-specified evaluation index \cite{bertsekas1995dynamic}, they require a central authority, leading to higher communication resource consumption and weaker system robustness. To avoid using the authority, decentralized strategies have been developed \cite{wang1973stabilization,corfmat1976decentralized,1429376,blanchini2014network}. Wang et al. \cite{wang1973stabilization} and Corfmat et al. \cite{corfmat1976decentralized} introduced a local control framework for stabilizing multi-channel systems. Later, some transmission constraints were considered under this framework \cite{1429376}. It is worth mentioning that decentralized strategies always confront structural constraints such that controllers may not exist, especially in the output-feedback control scheme. Although some structural linear matrix inequalities have been adopted to overcome this drawback \cite{blanchini2014network}, they are still restricted to many application scenarios. To fully bypass the structural issues, distributed strategies have been proposed by introducing communication between agents \cite{liu2017cooperative,liu2017cooperativeout,wang2020distributed,rego2021distributed,kim2020decentralized}. For example, by leveraging augmented observers in \cite{Martins}, Liu et al.   \cite{liu2017cooperative} and \cite{liu2017cooperativeout} introduced a joint state estimation and control framework to stabilize and regulate the output for a class of continuous multi-channel systems, respectively. Similar problems with communication constraints were investigated in \cite{rego2021distributed}. Recently, Kim et al. \cite{kim2020decentralized} realized the self-organization of multi-agent systems to regulate multi-channel systems by using Bass' algorithm.

However, a large amount of private information of agents needs to be shared with neighbors in the distributed control strategies, which poses many privacy issues \cite{lin2003multi}. In the literature, differential privacy \cite{mo2016privacy,yin2019accurate,kawano2020design} and state decomposition \cite{wang2019privacy,chen2020privacy,chen2021differential} are two typical techniques to handle these issues. However, these techniques mainly concentrate on consensus tasks and only preserve initial values of agents. The results on privacy preservation directly for cooperative control where agents behave differently and share information during the whole control process are still lacking.

\subsection{Motivations} \label{sec1.2}
The existing distributed cooperative control strategies have two unresolved points. First, in most works, such as \cite{liu2017cooperative,liu2017cooperativeout,wang2020distributed,rego2021distributed}, the state estimators and controllers of all agents are designed with global information, including all input and output matrices, albeit they can be locally performed. Therefore, the presence of a central authority remains necessary.  Second, since the shared messages among agents contain the private information, even when the estimators and controllers are designed in a distributed manner by utilizing Bass' algorithm, private information of agents may be leaked to or inferred by malicious neighbors under the framework in \cite{kim2020decentralized}. Also, the existing privacy-preserving mechanisms \cite{mo2016privacy,yin2019accurate,kawano2020design,wang2019privacy,chen2020privacy,chen2021differential} mainly preserve initial values in consensus tasks, they do not fit with the framework in \cite{kim2020decentralized}. The second point also reveals a significant technical dilemma: more information shared among agents will facilitate the design and the implement of strategies in a distributed manner and further benefit the self-organization of agents, but also bring more privacy issues. Till now, this dilemma has not been well solved.

Altogether, there remain two critical and urgent questions to be answered for the cooperative control problem of multi-channel systems: 1) how to design and perform a cooperative control strategy in a distributed way; 2) how to incorporating a privacy-preserving mechanism into the control strategy.

\subsection{Contributions} \label{sec1.3}
This paper studies the cooperative stabilization problem for a multi-channel system, where each channel is managed by an agent that can exchange information with its neighbors through a communication network. A novel cooperative control strategy is developed for each agent, and a state decomposition mechanism \cite{wang2019privacy} is embedded into the strategy to preserve all private information of each agent from neighbors' eavesdropping. Compared with the literature \cite{liu2017cooperative,liu2017cooperativeout,wang2020distributed,rego2021distributed,kim2020decentralized}, this paper possesses three advantageous features as follows:
 \begin{enumerate}
   \item Each agent possesses the capability to autonomously self-organize a privacy-preserving cooperative control strategy, allowing it to independently design and execute the strategy. ({\bf Theorems \ref{thm1}} and {\bf \ref{thm2}}). All private information can be preserved from adversaries' eavesdropping (Theorem \ref{privacythm}). 

   \item  The communication cost of the privacy-preserving mechanism is quantitatively analyzed {\bf (Theorem \ref{thm4}}). Further to reduce this cost, an optimization problem is introduced to select the mechanism parameter.

   \item Based on a linear quadratic regulator (LQR)-based index, the effect of adding more channels on the system behavior is analyzed in a closed form {({\bf Theorem \ref{thm5}})}, which provides a novel guide for channel placement.
 \end{enumerate}

\subsection{Organization and Notations} \label{sec1.4}

The rest of this paper is organized as follows. In Section \ref{sec2}, the system model and the problem statement are provided. In Section \ref{sec3}, a novel cooperative control strategy for multi-channel systems with self-organizing agents is designed, where a privacy-preserving mechanism is incorporated to preserve the private information of agents. In Section \ref{sec4}, some comparative simulation results are presented to illustrate the effectiveness of the theoretical results. In Section \ref{sec5}, a conclusion is drawn.

{
{\it{Notations}}: For a positive definite matrix  $S \in \mathbb{R}^{n \times n}$, $\lambda_{\text{min}}(S)$ and  $\lambda_{\text{max}}(S)$ denote its minimum and maximum eigenvalues, respectively. $I_n \in \mathbb{R}^{n \times n}$ is the identity matrix.  $0_n  \in \mathbb{R}^{n \times n}$ is a  matrix with all elements being $0$. $\mathbb{1}_n \in \mathbb{R}^{n }$ is a vector with all elements being $1$. Let $\mathbb{N}^{+}$ be the set of positive integers and $\mathbb{N} = \mathbb{N}^{+} \cup  0$.} 


















\section{Preliminaries and Problem Formulation}\label{sec2}

%
%

\subsection{System Model}\label{sec2.2}
This paper considers a multi-channel system, and each channel is managed by an agent. The dynamics of the multi-channel system are described by  
\begin{align} \label{equ:model_s}
  \begin{split}
s_{k+1} = & \ A s_{k} + \sum_{i=1}^{N} B^{i} u_{k}^i,   \\
  y_{k}^{i} = & \ C^i s_{k}, \ i \in \mathcal{V} \triangleq \{1,\ldots,N \}, \ k \in \mathbb{N},
\end{split}
\end{align}
where $N$ denotes the number of channels; the superscript $i$ denotes the $i$-th channel/agent; the subscript $k$ denotes the $k$-th time step; $s_{k} \in \mathbb{R}^{n}$ is the system state; $u_{k}^i \in \mathbb{R}^{r_i} $ is the system input generated by agent $i$; $y_{k}^i \in \mathbb{R}^{m_i} $ is the system output measured by agent $i$; and $A  \in \mathbb{R}^{n \times n} $, $B^{i} \in \mathbb{R}^{n \times r_i}  $, and $C^{i} \in \mathbb{R}^{m_i \times n}  $ are the system state, input and output matrices, respectively. Matrix $A$ is available for each agent. Before proceeding, denote $m = \sum_{i=1}^{N} m_{i}$ and $r = \sum_{i=1}^{N} r_{i}$. In this paper, each agent can measure a part of the state and exert a partial input on the system. Particularly, each agent has access to partial information of system (\ref{equ:model_s}), denoted by an information message $\mathcal{M}^i \triangleq \{ u_{k}^i$, $y_{k}^{i}$, $B^{i}$, $C^{i} \}$, and we call them  private agents. 
{
The communication network topology of agents is denoted by graph $\mathcal{G}$. The set of agent $i$'s neighbors is denoted by $\mathcal{N}_i$. We regard agent $i$ as its own neighbor, i.e., $i \in \mathcal{N}_i$, since it has access to its own information naturally. In addition, the row-stochastic matrix of graph $\mathcal{G}$ is defined as $W = [w_{ij}]_{N \times N} \in \mathbb{R}^{N \times N} $, where $w_{ij} > 0$ if $j \in \mathcal{N}_i$ but 0 otherwise, and $\sum_{j \in \mathcal{V}} w_{ij} = 1$.} We assume that the graph $\mathcal{G}$ is undirected and connected. For more knowledge about graph theory, please refer to \cite{ren2005consensus}.

\begin{assumption} \label{asm:stabilizability}   
The multi-channel system is jointly stabilizable, i.e., $(A, \ \mathcal{B})$ is stabilizable, where $\mathcal{B} = [B^1$, $\ldots$, $B^N] \in \mathbb{R}^{n \times r} $.
\end{assumption}

\begin{assumption}  \label{asm:detectability}   
The multi-channel system is jointly detectable, i.e., $(\mathcal{C}$, $A)$ is detectable, where $\mathcal{C}= [(C^1)^T$, $\ldots$, $(C^N)^T]^T \in \mathbb{R}^{m \times n} $.
\end{assumption}

%

\begin{lemma}  \cite{khan2011stability} \label{lemma1}
If Assumptions \ref{asm:stabilizability} and \ref{asm:detectability} hold,   $(A$, $\mathcal{B} \mathcal{B}^T)$ is stabilizable and $(\mathcal{C}^T \mathcal{C}$, $ A)$ is detectable, respectively.
\end{lemma}

{There may exist honest-but-curious adversarial agents in the multi-agent system \cite{wang2019privacy}. These adversaries know the network information $W$ and intend to learn neighbors' private information $\mathcal{M}^i$ based on the data received from their neighbors. Therefore, we assume that all agents will not exchange extra information with others, except for what is needed for the cooperative control strategy.

\begin{definition}  \label{definition1}  
The privacy of agent $i$ is preserved if adversaries cannot infer $\mathcal{M}^i$ with any guaranteed accuracy.   
\end{definition}
}

\subsection{Problem of Interest}\label{sec2.3}
This paper is aimed to design a cooperative control strategy for each agent to stabilize the multi-channel system in a distributed manner with privacy preserved. On the basis of the private information $\mathcal{M}^i$ and local interaction with neighbors, each agent estimates the full state of the multi-channel system (\ref{equ:model_s}) and utilizes the estimate to generate a local input to stabilize the multi-channel system in a cooperative way. In addition, the private information $\mathcal{M}^i$ is not inferred by its honest-but-curious adversarial neighbors.

{Specifically, we will design a distributed state estimator and a local controller for each agent $i$ at time step $k+1$, $\forall i \in \mathcal{V}$, $\forall k \in \mathbb{N}$, in the form of
\begin{align} \label{equ:taskform}
  \begin{split}
  z_{k+1}^i = & \ f(z_{k}^j, \ g(\mathcal{M}^j), \ j \in \mathcal{N}_i),    \\
  u_{k+1}^i = & \ h(z_{k+1}^{i}),  
  \end{split}
\end{align}
respectively, where $z_{k}^i$ is the estimate of $s_k$. According to (\ref{equ:taskform}), the cooperative task can be divided into two subtasks. On the one hand, we need to design an effective estimator and controller in forms of $f(\cdot)$ and $h(\cdot)$. On the other hand,  we need to develop an appropriate privacy-preserving mechanism $g(\cdot)$ to protect $\mathcal{M}^i$ against adversaries. 
}

\begin{remark}
The main challenges of addressing the above two subtasks lie in two aspects. First, the separation principle cannot be utilized as the state estimator and the controller are strongly coupled in the distributed framework. Therefore, they have to be jointly designed. Second, the preservation of $\mathcal{M}^i$  may restrict the information sharing among agents, which will further increase the difficulty of designing and implementing the designed state estimator and the controller. 
\end{remark}

\section{Main Results}\label{sec3}
In this section, a novel cooperative stabilization framework for the multi-channel system  (\ref{equ:model_s}) is proposed. In addition, a privacy-preserving mechanism is incorporated into the proposed framework with a thorough performance analysis.

\subsection{Joint State Estimator and Controller Design} \label{sec3.1}
In this subsection, a joint state estimation and control algorithm is designed for each agent to stabilize the  multi-channel system (\ref{equ:model_s}), based only on its private message $\mathcal{M}^i$ and local interaction with neighbors. First of all, the state estimation of the multi-channel system  (\ref{equ:model_s}) is considered. The estimate of $s_{k+1}$ for agent $i$ is denoted by $z_{k+1}^i$, which is designed as
\begin{equation} \label{equ:model_zi}
  \left \{ \begin{array}{l}
z_{k+1}^i =     x_{k+1,M_1}^i,   \\
 x_{k+1,l}^i =   \sum_{j \in \mathcal{N}_i} w_{ij} x_{k+1,l-1}^j,  \\
 x_{k+1,0}^i =   A z_{k}^i +  N B^{i} u_{k}^i + N L^{i}(y_{k}^{i} - C^{i} z_{k}^i),
  \end{array} \right .
\end{equation}
where $ l = 1$, $\ldots$, $M_1$; $M_1$ is the fusion step to be designed later; $w_{ij}$ is the coupling gain defined in Section \ref{sec2.2}; and $L^{i}$ is the estimator gain that makes $A - \sum_{i=1}^{N}  L^{i} C^{i} $ Schur stable. Then, the cooperative controller for stabilizing the multi-channel system  (\ref{equ:model_s}) is proposed. For agent $i$, the input $u_{k}^i$ is designed as
\begin{align} \label{equ:model_ui}
u_{k}^i =   & \ K^i z_{k}^i,
\end{align}
where  $K^{i}$ is the control gain ensuring that $A + \sum_{i=1}^{N}  B^{i} K^{i} $ is Schur stable. To proceed, {denote  $A_1 = A + \sum_{i=1}^{N}  B^{i} K^{i}  $ and $A_2 = A - \sum_{i=1}^{N}  L^{i} C^{i}$}. Since both $A_1$ and $A_2$ are Schur stable,
the matrix
\begin{align}  \label{equ:F}
F = \left[{
  \begin{array}{c  c}
  {A_1} & { \sum_{i=1}^{N}  B^{i} K^{i} }   \\ 
  { 0 } & {A_2}  \\
  \end{array}} \right]
\end{align}
is also Schur stable. {Hence, for any positive definite matrix $Q_0$, there exists a positive definite matrix $P_0$ such that the following Lyapunov equation holds \cite[Theorem 5.D5]{chen1999linear}}
\begin{align} 
F^T P_0 F -   P_0 + Q_0 =   0. \notag
\end{align}
Note that there always exists a sufficiently small positive scalar $\theta $ such that $P = \frac{1}{1+\theta} P_0 > 0$ and $Q = Q_0 - \frac{\theta}{1+\theta} P_0 >0$. Then, the above equality can be rewritten as
\begin{align}  \label{equ:lyapunove}
(1 +  \theta) F^T P F -   P + Q = 0.
\end{align}
Besides, define an auxiliary constant matrix as
\begin{align} \label{equ:mathcalA}
\mathcal{A}  = I_N \otimes A -  N \bar{\mathcal{L}} \bar{\mathcal{C}} +  N \bar{\mathcal{B}} \bar{\mathcal{K}},
\end{align}
where $\bar{\mathcal{L}} = \text{diag}\{L^1$, $\ldots$, $L^N\}$, $\bar{\mathcal{C}} = \text{diag}\{C^1$, $\ldots$, $C^N\}$, $\bar{\mathcal{B}} = \text{diag}\{B^1$, $\ldots$, $B^N\}$, and $\bar{\mathcal{K}} = \text{diag}\{K^1$, $\ldots$, $K^N\}$.

\begin{theorem} \label{thm1}
Suppose Assumptions \ref{asm:stabilizability} and \ref{asm:detectability} hold. Then, the cooperative control strategy (\ref{equ:model_zi}) and (\ref{equ:model_ui}) stabilizes the multi-channel system (\ref{equ:model_s}) if
\begin{align} \label{equ:M1}
M_1 > \bar M_1 \triangleq \frac{1}{2} \log_{\lambda} \bigg \{    \frac{\theta^2  }{2 (1+ \theta)^2 \psi^2  } \frac{ \lambda_{\text{min}}(Q)}{  \lambda_{\text{max}}(P) }   \bigg \},
\end{align}
where $\lambda $ is the second largest eigenvalue of $W$, matrices $Q$ and $P$ are given in (\ref{equ:lyapunove}), and $\psi =   \max \Big \{ 1$, $\|  N \bar{\mathcal{B}}   \bar{\mathcal{K}} \|_2^2 +  \| \mathcal{A} -  N \bar{ \mathcal{B}}   \bar{\mathcal{K}} \|_2^2$, $\|\mathcal{A} + N \bar{\mathcal{L}}  \bar{\mathcal{C}} \|_2^2 + \| \mathcal{A} \|_2^2 \Big \}$.
\end{theorem}

The proof of Theorem \ref{thm1} is provided in Appendix \ref{thm1proof}. {Note that $\bar M_1$ in (\ref{equ:M1}) is computed using the system matrices of all agents, which indicates that the proposed cooperative strategy (\ref{equ:model_zi}) and (\ref{equ:model_ui}) depends on global information. To deal with this issue, we can calculate $\bar M_1$ in (\ref{equ:M1}) using the bounds of system matrices. As shown in Remark \ref{remarkdistributed}, the bounds of system matrices can be obtained by each agent in a distributed way. The cooperative control strategy can be designed without any central authority.  
}

\begin{assumption} \label{asm:boundness}   
There exist positive scalars $\kappa_A$, $\kappa_B$,  $\kappa_C$, $\kappa_K$, $\kappa_L$, $\kappa_P$ and $\kappa_{Q}$ satisfying $\| A \|_2 \leq \kappa_A $, $\| B^i \|_2  \leq \kappa_B $, $\| C^i \|_2  \leq \kappa_C $, $\| K^i \|_2  \leq \kappa_K $, $\| L^i \|_2  \leq \kappa_L $, $\| P \|_2  \leq \kappa_P $, and $ \kappa_{Q} I_{2n} \leq   Q $, $\forall i \in \mathcal{V}$.
\end{assumption}

{
\begin{corollary} \label{cor1}
If Assumption \ref{asm:boundness} holds, then the condition (\ref{equ:M1}) can be simplified by
\begin{align} \label{equ:M12}
M_1 >   \log_{\lambda} \bigg \{   \frac{  \sqrt{2} \theta    }{ 2 (1+ \theta)   \psi_1 }  \sqrt \frac{\kappa_{Q}}{\kappa_P}\bigg \},
\end{align}
where $ \psi_1 = \max \Big \{ 1, \  5 \kappa_A^2 +   5 N^2  \kappa_B^2 \kappa_K^2 +  3 N^2  \kappa_L^2 \kappa_C^2  \Big \}$.
\end{corollary}
}
  
The proof of Corollary \ref{cor1} is provided in Appendix \ref{cor1proof}. {By utilizing the boundedness of system matrices, we provide another way (\ref{equ:M12}) to design the fusion parameter $M_1$. This approach does not undermine the privacy-preserving performance to be considered.}  
 
%


\subsection{Privacy-Preserving State Estimator and Controller Design}\label{sec3.2}
In this subsection, a privacy-preserving mechanism is incorporated into the joint state estimator (\ref{equ:model_zi}) and controller (\ref{equ:model_ui}) to preserve the private information $\mathcal{M}^i$ for each agent.
First, we introduce two auxiliary vectors $z_{k+1,\alpha}^{i}$ and $z_{k+1,\beta}^{i}$ 
\begin{align} \label{equ:model_zalphai}
  \left \{ \begin{array}{l} \! 
z_{k+1,\alpha}^{i} \!  =  \! \alpha_{k+1,M_2}^{i}, \qquad  z_{k+1,\beta}^{i} =  \beta_{k+1,M_2}^{i},   \\
  \alpha_{k+1,l}^i \!  = \!   \sum_{j \in \mathcal{N}_i} w_{ij} \alpha_{k+1,l-1}^j \!  - \!   \epsilon \pi_i (\alpha_{k+1,l-1}^i \!  - \beta_{k+1,l-1}^i  ),  \\
  \beta_{k+1,l}^i \!  = \!  (1 - \epsilon\pi_i ) \beta_{k+1,l-1}^i \!  + \! \epsilon\pi_i \alpha_{k+1,l-1}^i,  l=1, \ldots, M_2, \\
\end{array} \right .
\end{align}
where $M_2$ is the fusion step to be designed later; $\epsilon \in (0$, $\frac{2}{3})$ is a scalar; {$\pi_i \in (0, \ 1)$ is a private scalar generated and owned only by agent $i$;} other parameters are defined the same as those in Section \ref{sec3.1};  $\alpha_{k+1,0}^i =    \pi_i [ A z_{k}^i +  N B^{i} u_{k}^i + N L^{i}(y_{k}^{i} - C^{i} z_{k}^i) ]$, $\beta_{k+1,0}^i =  (1- \pi_i) [ A z_{k}^i +  N B^{i} u_{k}^i + N L^{i}(y_{k}^{i} - C^{i} z_{k}^i)]$. Then, the estimate of $s_{k+1}$ for agent $i$ is designed as
\begin{align} \label{equ:model_z2i}
z_{k+1}^{i} =   & \ z_{k+1,\alpha}^{i} +  z_{k+1,\beta}^{i}.
\end{align}

\begin{theorem} \label{thm2}
Suppose Assumptions \ref{asm:stabilizability} and \ref{asm:detectability} hold. Then, the cooperative control strategy (\ref{equ:model_z2i}) and (\ref{equ:model_ui}) stabilizes the multi-channel system (\ref{equ:model_s}) if
\begin{align} \label{equ:M2}
M_2 >   \bar M_2 \triangleq  \frac{1}{2} \log_{\tilde{\lambda}} \bigg \{    \frac{\theta^2  }{2 (1+ \theta)^2 \tilde{\psi}^2  } \frac{ \lambda_{\text{min}}(Q)}{  \lambda_{\text{max}}(P) }   \bigg \},
\end{align}
where $\tilde{\lambda} $ is the second largest absolute value of eigenvalues of $\tilde{W}$, $\tilde{\psi} =   \max \Big \{ 1$, $\|  N \bar{\mathcal{B}}   \bar{\mathcal{K}} \|_2^2 +  \| \mathcal{A} -  N \bar{ \mathcal{B}}   \bar{\mathcal{K}} \|_2^2$, $2 \| V \|_2^2 ( \|\mathcal{A} + N \bar{\mathcal{L}}  \bar{\mathcal{C}} \|_2^2 + \| \mathcal{A} \|_2^2 ) \Big \}$, $\Pi = \text{diag}\{\pi_1$, $\ldots$, $\pi_N\}$, and 
\begin{align} 
V =  &  \left [
  \begin{array}{c c}
  { \Pi  }   \\
  { I_{N} -   \Pi} \end{array}  \right ], \
  \tilde{W} = \! & \! \left [
  \begin{array}{c c}
  {W  - \epsilon \Pi } & {\epsilon \Pi  }  \\
  {  \epsilon \Pi   } & {  (1 - \epsilon) \Pi  } \end{array}  \right  ]. \notag 
\end{align}
\end{theorem}

The proof of Theorem \ref{thm2} is provided in Appendix \ref{thm2proof}. To remove the global information $V$, $\tilde{\psi}$ in (\ref{equ:M2}) can be replaced by $ \tilde{\psi}_1 =   \max \Big \{ 1, \  \|  N \bar{\mathcal{B}}   \bar{\mathcal{K}} \|_2^2 +  \| \mathcal{A} -  N \bar{ \mathcal{B}}   \bar{\mathcal{K}} \|_2^2$, $2 ( \|\mathcal{A} + N \bar{\mathcal{L}}  \bar{\mathcal{C}} \|_2^2 + \| \mathcal{A} \|_2^2 ) \Big \}$ since $\| V \|_2 \in (\frac{1}{2}$, $1)$. Then, by combining the proofs of Corollary \ref{cor1} and Theorem \ref{thm2}, if Assumption \ref{asm:boundness} holds, the condition (\ref{equ:M2}) can be simplified as
  \begin{align}  \label{equ:M22}
  M_2 >   \log_{\tilde{\lambda}} \bigg \{   \frac{  \sqrt{2} \theta    }{ 2 (1+ \theta)   \psi_2 }  \sqrt \frac{\kappa_{Q}}{\kappa_P}\bigg \}, 
  \end{align}
  where $ \psi_2 = \max \Big \{ 1, \  10 \kappa_A^2 +   10 N^2  \kappa_B^2 \kappa_K^2 +  6 N^2  \kappa_L^2 \kappa_C^2  \Big \}$.


In the above design, the state estimator gain $L^i$ and the controller gain $K^i$ are subject to global constraints, i.e., $A - \sum_{i=1}^{N}  L^{i} C^{i} $ and $A + \sum_{i=1}^{N}  B^{i} K^{i} $ should be Schur stable. In the following, a distributed method to design these gains is proposed with privacy preserved. First, we consider the design of $K^i$ for agent $i$, $i \in \mathcal{V}$, in a fully distributed manner. To proceed, we introduce an auxiliary matrix $\mathcal{B}^{i}$, defined as
 \begin{equation} \label{equ:model_Bi}
  \left \{ \begin{array}{l}
 \mathcal{B}^{i} =  \bar{\mathcal{B}}^{i}_{M_3} + \hat{\mathcal{B}}^{i}_{M_3},    \\
 \bar{\mathcal{B}}^{i}_{h} =    \sum_{j \in \mathcal{N}_i} w_{ij} \bar{\mathcal{B}}^{j}_{h-1} - \epsilon \pi_i (\bar{\mathcal{B}}^{j}_{h-1} - \hat{\mathcal{B}}^{j}_{h-1}),  \\
 \hat{\mathcal{B}}^{i}_{h} = (1 - \epsilon \pi_i ) \hat{\mathcal{B}}^{j}_{h-1} + \epsilon \pi_i \bar{\mathcal{B}}^{j}_{h-1}, \ h = 1,\ldots, M_3,
  \end{array} \right .
\end{equation}
where $\bar{\mathcal{B}}^{i}_{0} = \pi_i N B^{i} (B^{i})^T$, $\hat{\mathcal{B}}^{i}_{0} = (1- \pi_i) N B^{i} (B^{i})^T$, and $\epsilon$ and $\pi_i$ are defined below (\ref{equ:model_zalphai}). Then, $K^i$ is selected as
\begin{align} \label{equ:kd}
K^i = - ( B^{i})^T [ I_{n} + (\mathcal{B}^{i})^T P_K^i \mathcal{B}^{i}]^{-1} (\mathcal{B}^{i})^T P_K^i A,
\end{align}
where $P_K^i $ satisfies
\begin{align} 
 P_K^i =  & I_n  -  A^T   P_K^i \mathcal{B}^{i} [(\mathcal{B}^{i})^T P_K^i \mathcal{B}^{i} + I_{n}]^{-1} (\mathcal{B}^{i})^T  P_K^i A \notag \\
 & + A^T   P_K^i  A. \notag
 \end{align}
Similarly, the state estimator gain $L^i$ is designed as
\begin{align} \label{equ:ld}
L^i = A P_L^i (\mathcal{C}^{i})^T [I_{n} +  \mathcal{C}^{i} P_L^i (\mathcal{C}^{i})^T]^{-1}  (C ^{i})^T,
\end{align}
where $\mathcal{C}^{i}$ is derived by
 \begin{align} \label{equ:model_Ci}
  \left \{ \begin{array}{l}
 \mathcal{C}^{i} =   \bar{\mathcal{C}}^{i}_{M_4} + \hat{\mathcal{C}}^{i}_{M_4},    \\
 \bar{\mathcal{C}}^{i}_{h} =  \sum_{j \in \mathcal{N}_i} w_{ij} \bar{\mathcal{C}}^{j}_{h-1} - \epsilon \pi_i (\bar{\mathcal{C}}^{j}_{h-1} - \hat{\mathcal{C}}^{j}_{h-1}),  \\
 \hat{\mathcal{C}}^{i}_{h} = (1 - \epsilon  \pi_i ) \hat{\mathcal{C}}^{j}_{h-1} + \epsilon \pi_i \bar{\mathcal{C}}^{j}_{h-1}, \ h = 1,\ldots, M_4,
\end{array} \right .
\end{align}
with $\bar{\mathcal{C}}^{i}_{0} = \pi_i N (C^{i})^T  C^{i}$, $\hat{\mathcal{C}}^{i}_{0} = (1- \pi_i) N  (C^{i})^T C^{i}$, and  
\begin{align} 
P_L^i =  & I_n -  A   P_L^i (\mathcal{C}^{i})^T   [  \mathcal{C}^{i} P_L^i (\mathcal{C}^{i})^T + I_{n}]^{-1} \mathcal{C}^{i}  P_L^i A^T  \notag \\
 & + A   P_L^i  A^T . \notag 
\end{align}

\begin{lemma} \label{thm3}
Suppose that Assumptions \ref{asm:stabilizability} and \ref{asm:detectability} hold. $\mathcal{B}^{i}$ and $\mathcal{C}^{i}$ in (\ref{equ:model_Bi}) and (\ref{equ:model_Ci}) converge to $\mathcal{B} \mathcal{B}^T$ and $\mathcal{C}^T \mathcal{C}$ exponentially with respect to $M_3$ and $M_4$, respectively. When $M_3$ and $M_4$ tend to infinity, $K^i$ and $L^i$ designed in (\ref{equ:kd}) and (\ref{equ:ld}) ensure that $A + \sum_{i=1}^{N} B^{i} K^{i} $ and $A - \sum_{i=1}^{N}  L^{i} C^{i} $ are Schur stable, respectively.
\end{lemma}

{ The proof of Lemma \ref{thm3} is provided in Appendix \ref{thm3proof}.} The proposed privacy-preserving cooperative control strategy for agent $i$ is summarized as Algorithm \ref{algorithm1}. Note that it is impossible to set $M_3 = \infty$ and $M_4 = \infty$ in practice. To address this issue, taking $M_3$ for an example, we introduce a terminal condition in Algorithm \ref{algorithm1}, i.e., $\|\bar{\mathcal{B}}^{i}_{h} - \bar{\mathcal{B}}^{i}_{h-1}\|_2 \leq \delta$ where $\delta$ is a small positive constant. Since $\bar{\mathcal{B}}^{i}_{h}$ converges to a fixed value exponentially as $\mathcal{B}^{i}$, this condition with $\delta=0.001$ is effective as shown in the simulation part. 

\begin{remark}  \label{remarkdistributed}
  {In the existing works \cite{liu2017cooperative,liu2017cooperativeout,wang2020distributed,rego2021distributed}, $L^i$ and $K^i$ are designed using a central authority to collect all system input and output matrices $C^{i}$ and $ B^{i}$.} In this paper, it follows from (\ref{equ:model_Bi})-(\ref{equ:model_Ci}) that they are selected without utilizing the authority. {In addition, the bounds of system matrices in Assumption \ref{asm:boundness} can also be computed based on (\ref{equ:model_Bi})-(\ref{equ:model_Ci}) for each agent in a distributed way. Take the bound $\kappa_B$ for an example. First, each agent can directly obtain the value of $\mathcal{B}  \mathcal{B}^T$ using (\ref{equ:model_Bi}). Then, noting that $\mathcal{B}  \mathcal{B}^T \ge B^i (B^i)^T$, $\forall i \in \mathcal{V}$, we can let $\kappa_B = \sqrt{ \| \mathcal{B}  \mathcal{B}^T \|_2} $ be the bound since $\| B^i \|_2  \leq \kappa_B $, $\forall i \in \mathcal{V}$. Other  bounds in Assumption \ref{asm:boundness} can be similarly derived by using the fusion law (\ref{equ:model_Bi})-(\ref{equ:model_Ci}).} Altogether, both the design and the implementation of the cooperative control strategy can be carried out by each agent. This feature makes it feasible to introduce additional channel/agent into the system, which will not affect most of the other agents globally. Thus, the proposed control strategy possesses a better scalability.
\end{remark}

\begin{algorithm}[t]
\caption{Privacy-Preserving Control Strategy for Agent $i$}
\hspace*{0.02in}

{\bf Gains predesign:}

\label{algorithm1}
\begin{algorithmic}[1]

\State  initialize $\pi_i$; $h \leftarrow 1 $; $\delta$;

\State  {\bf repeat}

\State  exchange $\bar{\mathcal{B}}^{i}_{h-1}$ with neighbors;

\State  compute $\bar{\mathcal{B}}^{i}_{h}$ and $\hat{\mathcal{B}}^{i}_{h}$ by (\ref{equ:model_Bi});

\State    $ h \leftarrow h + 1$;

\State  {\bf until} $\|\bar{\mathcal{B}}^{i}_{h} - \bar{\mathcal{B}}^{i}_{h-1}\|_2 \leq \delta$

\State  $ \mathcal{B}^{i} \leftarrow   \  \bar{\mathcal{B}}^{i}_{h} + \hat{\mathcal{B}}^{i}_{h}$ and obtain
 $ \mathcal{C}^{i}$ similarly;

\State compute $K^i$ and $L^i$ by (\ref{equ:kd}) and  (\ref{equ:ld}), respectively.

\State {\bf output:} $K^i$ and $L^i$;

\end{algorithmic}

{\bf Online cooperative control at time step $k$: }

\begin{algorithmic}[1]



\State  initialize $\alpha_{k,0}^{i}$ and $\beta_{k,0}^{i}$ by (\ref{equ:model_zalphai}), respectively;

\State  {\bf for} $l=1:M_2$

\State  exchange $\alpha_{k,l-1}^{i}$ with neighbors;

\State  compute $\alpha_{k,l}^{i}$ and $\beta_{k,l}^{i}$ by (\ref{equ:model_zalphai}), respectively;

\State  {\bf end}

\State $ z_{k}^{i} =   \alpha_{k,M_2}^{i}  +  \beta_{k,M_2}^{i}$;

\State exert $u_{k}^i = \ K^i z_{k}^i$ on the multi-channel system;



\end{algorithmic}
\end{algorithm}

{
According to the definition of adversaries in Section \ref{sec2.2}, they attempt to infer the private information $\mathcal{M}^i$ of neighbors based on the exchanged data. Without loss of generality, let agent $i$ have such an adversarial neighbor denoted by adversary $\zeta$, $\zeta \in \mathcal{N}_i$. It follows from Algorithm \ref{algorithm1} that all information of agent $i$ exchanged with adversary $\zeta$ up to time step $k+1$ can be denoted by $I_{i \zeta} = \{\bar{\mathcal{B}}^{i}_{h}$, $\bar{\mathcal{C}}^{i}_{h}$, $\alpha_{t,l}^{i}\}_{h=0, l=0, t=0}^{M_3-1, M_2-1, k}$.} Based on these notations, an important privacy-preserving result of Algorithm \ref{algorithm1} can be obtained as follows.

{
\begin{theorem} \label{privacythm}
  Under Algorithm \ref{algorithm1}, adversary $\zeta$ cannot infer $\mathcal{M}^i$ with any guaranteed accuracy if $\mathcal{N}_i \not \subset \mathcal{N}_{\zeta}$. 
  \end{theorem}
}

The proof of Theorem \ref{privacythm} is provided in Appendix \ref{privacythmproof}. {The state decomposition technique was introduced for consensus in \cite{wang2019privacy}. We apply this technique in (\ref{equ:model_zalphai}), (\ref{equ:model_Bi}) and (\ref{equ:model_Ci}) for achieving the goal of preserving privacy. We take the one in (\ref{equ:model_zalphai}) for an example to demonstrate the basic idea. Specifically, in the original state estimator (\ref{equ:model_zi}) without state decomposition, the state estimate $x_{k,l}^i$ containing the private information of agent $i$ is directly transmitted to neighbors. Subsequently, the curious neighbor may infer the private information of agent $i$ based on the received data. To address this issue, the state estimate $x_{k,l}^i$ is decomposed into two parts $\alpha_{k,l}^i$ and $\beta_{k,l}^i $ in (\ref{equ:model_zalphai}) using the private parameter $\pi_i$. Then, agent $i$ only shares $\alpha_{k,l}^i$ with its neighbors at each information sharing step. Since the curious neighbor does not know the value of $\pi_i$ or $\beta_{k,l}^i$, it cannot recover the value of $x_{k,l}^i$.}  We apply it to the cooperative control of multi-channel systems. In addition, this paper provides a mathematical view to explain why adversaries cannot recover the private information, i.e., the variables for adversaries to recover the private information are always more than the constraints they have due to the introduction of $\pi_i$.

{
\subsection{Robustness Analysis of Algorithm \ref{algorithm1}} \label{sec3robust}
This part shows that Algorithm \ref{algorithm1} can still guarantee the boundedness of system (\ref{equ:model_s}) with both the process and measurement noise. Particularly, the noisy model is described by
\begin{align} \label{equ:model_noise}
  \begin{split}
s_{k+1} = & \ A s_{k} + \sum_{i=1}^{N} B^{i} u_{k}^i + \omega_k ,   \\
  y_{k}^{i} = & \ C^i s_{k} + \nu_k^i,
  \ i \in \mathcal{V}, \ k \in \mathbb{N}, 
\end{split}
\end{align}
where $\omega_k \sim  \mathcal{N}(0,\ \sigma_{\omega}^2 I_n) $ and $\nu_k^i   \sim  \mathcal{N}(0,\ \sigma_{\nu}^2 I_{m_i}) $ are zero-mean process and  measurement Gaussian noise, respectively, while the remaining variables are defined the same as those in (\ref{equ:model_s}).} Let $\sigma_{\omega}^2 I_n > 0$ and $\sigma_{\nu}^2 I_{m_i} > 0$ denote the noise covariances. We assume that $\omega_{k}$ and $\nu_{k}^i$, $\forall k \in \mathbb{N}$, $\forall i \in \mathcal{V}$, are mutually uncorrelated. 
Before moving on, a result about $F_2$ defined below (\ref{equ:eta2}) is given as a preliminary. According to \cite[Theorem 5.D5]{chen1999linear}, since $F_2$ is Schur stable, there always exist two positive definite matrices $\breve{P}$ and $\breve{Q}$ that  
\begin{align}  \label{equ:lyapunovbreve}
  F_2^T \breve{P} F_2 -  \breve{P} + \breve{Q} = 0,
  \end{align}
and $\breve{\theta} \triangleq 1 -  \lambda_{\text{min}} (\breve{Q} )/\lambda_{\text{max}} ( \breve{P} ) \in (0$, $1)$.

{
\begin{corollary} \label{corollarynoise} 
  Algorithm \ref{algorithm1} guarantees that the ultimate state of the noisy system (\ref{equ:model_noise}) is bounded in the form of  
  \begin{align} \label{equ:sbound}
    \lim_{k \rightarrow \infty} \mathbb{E} \{ \| s_k\|_2^2 \} \leq    \frac{ \lambda_{\text{max}} (F_\omega^T \breve{P} F_\omega )( n \sigma_{\omega}^2 + m \sigma_{\nu}^2 )}{ \lambda_{\text{min}} ( \breve{P} )(1 - \breve{\theta}) }, 
  \end{align}
  if conditions in Theorem \ref{thm2} hold, where $\breve{P}$ and $\breve{\theta}$ are given in (\ref{equ:lyapunovbreve}), and $
    F_\omega = [I_n$, $(N ( \tilde{W}^{M_2} V \otimes I_n)  \bar{\mathcal{L}})^T]^T$.  
\end{corollary} 
}
 
The proof of Corollary \ref{corollarynoise} is provided in Appendix \ref{corollarynoiseproof}. In the following, the estimation performance of the proposed estimator (\ref{equ:model_zalphai}) in Algorithm \ref{algorithm1} for the noisy system (\ref{equ:model_noise}) is analyzed. {Let $e_{k}^i = z_{k,M_2}^i - s_k$ be the estimation error of agent $i$ at step $k$, and $e_{k} =[(e_{k}^1)^T$, $\ldots$, $(e_{k}^N)^T]^T$ be the augmented estimation error.} Introduce the auxiliary matrix $I_e =[-\mathbb{1}_N \otimes I_n$, $I_{2Nn}$, $I_{2Nn}]$. By referring to $\breve{\eta}_k $ defined in the proof of Corollary \ref{corollarynoise}, we have $e_k = I_e \breve{\eta}_k $ and 
$$
\mathbb{E} \{ e_{k} e_{k}^T \} \leq  \mathbb{E} \{ I_e \breve{\eta}_{k} \breve{\eta}_{k}^T I_e^T\} \leq \|I_e\|^2_2 \mathbb{E} \{ \breve{\eta}_{k} \breve{\eta}_{k}^T\}.
$$ 
Further, by adopting the derivation techniques in the proof of Corollary \ref{corollarynoise}, it can be similarly derived that $ \mathbb{E} \{ \breve{\eta}_{k} \breve{\eta}_{k}^T \} \leq P_{e,k} $, where 
$$
{ P_{e,k} = F_2 P_{e,k-1} F_2^T + F_{\omega} Q_{\omega} F_\omega^T,} 
$$ 
with $Q_{\omega} = \text{diag} \{ \sigma_{\omega}^2 I_n$, $ \sigma_{\nu}^2  I_m\}$. Particularly, since $F_2$ is Schur stable, $P_{e,k}$ converges to $P_{e}$, which is the unique solution to $P_{e} = F_2 P_{e} F_2^T + F_{\omega} Q_{\omega} F_\omega^T $. {Altogether, we have
$$\mathbb{E} \{ e_{k} e_{k}^T \} \leq \|I_e\|^2_2 P_{e,k}, \ k \in \mathbb{N},  $$
and 
$$\lim_{k \rightarrow \infty } \mathbb{E} \{ e_{k} e_{k}^T \} \leq \|I_e\|^2_2 P_{e},  $$
which provides an upper bound for the covariance of estimation error. Moreover, if no system noise exists, i.e., $Q_{\omega}=0$, we have $ \lim_{k \rightarrow \infty }\mathbb{E} \{ e_{k} e_{k}^T \} = 0$. 
}

\subsection{Cost Analysis of the Privacy-Preserving Mechanism}\label{sec3.4}
In this subsection, the cost of adding the proposed privacy-preserving mechanism is analyzed. Specifically, the convergence rates between algorithms with and without privacy preservation are compared.
To be quantitative, $\bar M_1$ and $\bar  M_2$ in Theorems \ref{thm1} and \ref{thm2} are compared.

\begin{theorem} \label{thm4}
 $\bar M_1 < \bar  M_2$,  $\forall \epsilon \in (0, \ \frac{2}{3})$,  $\forall \pi_i \in (0, 1)$.
\end{theorem}

The proof of Theorem \ref{thm4} is provided in Appendix \ref{thm4proof}.
To guarantee the system stability, the proposed control strategies with and without the privacy-preserving mechanism require $M_1$ and $M_2$ fusion steps during each control interval, respectively. Theorem \ref{thm4} reveals that a higher communication overhead is needed to realize privacy preservation, which coincides with many existing privacy-preserving methods \cite{mo2016privacy,wang2019privacy}. In addition, to achieve a better trade-off between privacy preservation and communication overhead, the parameter $\epsilon$ in the  privacy-preserving mechanism can be selected by solving an optimization problem:
\begin{align} 
  \min_{\epsilon}  \max_{\lambda_i} |  1+ \lambda_i \pm \sqrt{(1 - \lambda_i )^2 + 4 \epsilon^2 }  - 2 \epsilon|, \notag
\end{align}
where $\lambda_i$ represents the eigenvalue of $W$ expect 1.  According to the proof of Theorem \ref{thm4}, when the above optimization problem is solved, $\tilde{\lambda}_{i}$ is minimized and thus the value of $M_2$ is minimized. Hence, the communication cost of the proposed privacy-preserving algorithm can be reduced.

\subsection{Performance Analysis of Adding More Channels}\label{sec3.5}

In this subsection, an LQR-based performance of system (\ref{equ:model_s}) under the proposed cooperative control strategy is introduced and the effect of adding  more channels on this performance is analyzed.
First, introduce a normalized LQR index widely used in the literature \cite{bertsekas1995dynamic,zhou1996robust} as  
\begin{align}  \label{equ:lqrindex}
J = s_L^T s_L +  \sum_{k=0}^{L-1} ( s_k^T s_k + u_k^T u_k ), \quad L \rightarrow \infty.
\end{align}

\begin{proposition} \label{pro1}
For the multi-channel system (\ref{equ:model_s}), suppose Assumptions \ref{asm:stabilizability} and \ref{asm:detectability} hold. Algorithm \ref{algorithm1} ensures the performance index in (\ref{equ:lqrindex}) is minimized as $J_0 = s_0^T P_0 s_0 $ when $M_2$ tends to infinity, where
\begin{align}
 P_0  =  & I_n + A^T   P_0  A   -  A^T   P_0 \mathcal{B}_0 ( \mathcal{B}_0^T P_0 \mathcal{B}_0  + I_{n})^{-1} \mathcal{B}_0^T P_0 A  \notag
 \end{align}
 with $\mathcal{B}_0 = \mathcal{B} \mathcal{B}^T$.
\end{proposition}

Proposition \ref{pro1} can be directly obtained by referring to \cite[Chapter 13]{zhou1996robust}. Without loss of generality, we assume that one more channel is added, denoted by channel $N+1$ managed by agent $N+1$, into the system (\ref{equ:model_s}). The new system dynamics can be described as
\begin{align} \label{equ:model_snew}
s_{k+1} = & \ A s_{k} + \sum_{i=1}^{N+1} B^{i} u_{k}^i,
\end{align}
where $B^{i}$, $i \in \mathcal{V}$, remains unchanged; $ B^{N+1} \in \{ e^{1}$, $\ldots$, $e^{n} \} $ and $e^{j} \in \mathbb{R}^n$ is a vector with the $j$-th element being $1$ and other elements being $0$. Just for presentation simplicity, $ B^{N+1} $ is regarded as a vector. For system (\ref{equ:model_snew}), the performance index $J$ can be minimized as $J_1 = s_0^T P_1 s_0 $ by utilizing Algorithm \ref{algorithm1}, where
\begin{align}
 P_1  =  & I_n + A^T   P_1  A   -  A^T   P_1 \mathcal{B}_1 ( \mathcal{B}_1^T P_1 \mathcal{B}_1  + I_{n})^{-1} \mathcal{B}_1^T P_1 A \notag
 \end{align}
with $\mathcal{B}_1 = \mathcal{B} \mathcal{B}^T + B^{N+1} (B^{N+1})^T $.

\begin{theorem} \label{thm5}
  $J_0 \ge J_1$.
\end{theorem}

The proof of Theorem \ref{thm5} is provided in Appendix \ref{thm5proof}. Theorem \ref{thm5} reveals that adding more channels can improve the LQR performance under the proposed control framework when the communication resource is sufficient. This feature provides a preliminary guide for channel placement. For example, in the application scenarios with high communication budget and performance requirement, it is suggested to arrange more channels/agents to participate in the cooperation task.


\section{Simulation}\label{sec4}
\begin{figure}[!htb]
\center
\subfigure{{\includegraphics[scale=0.33]{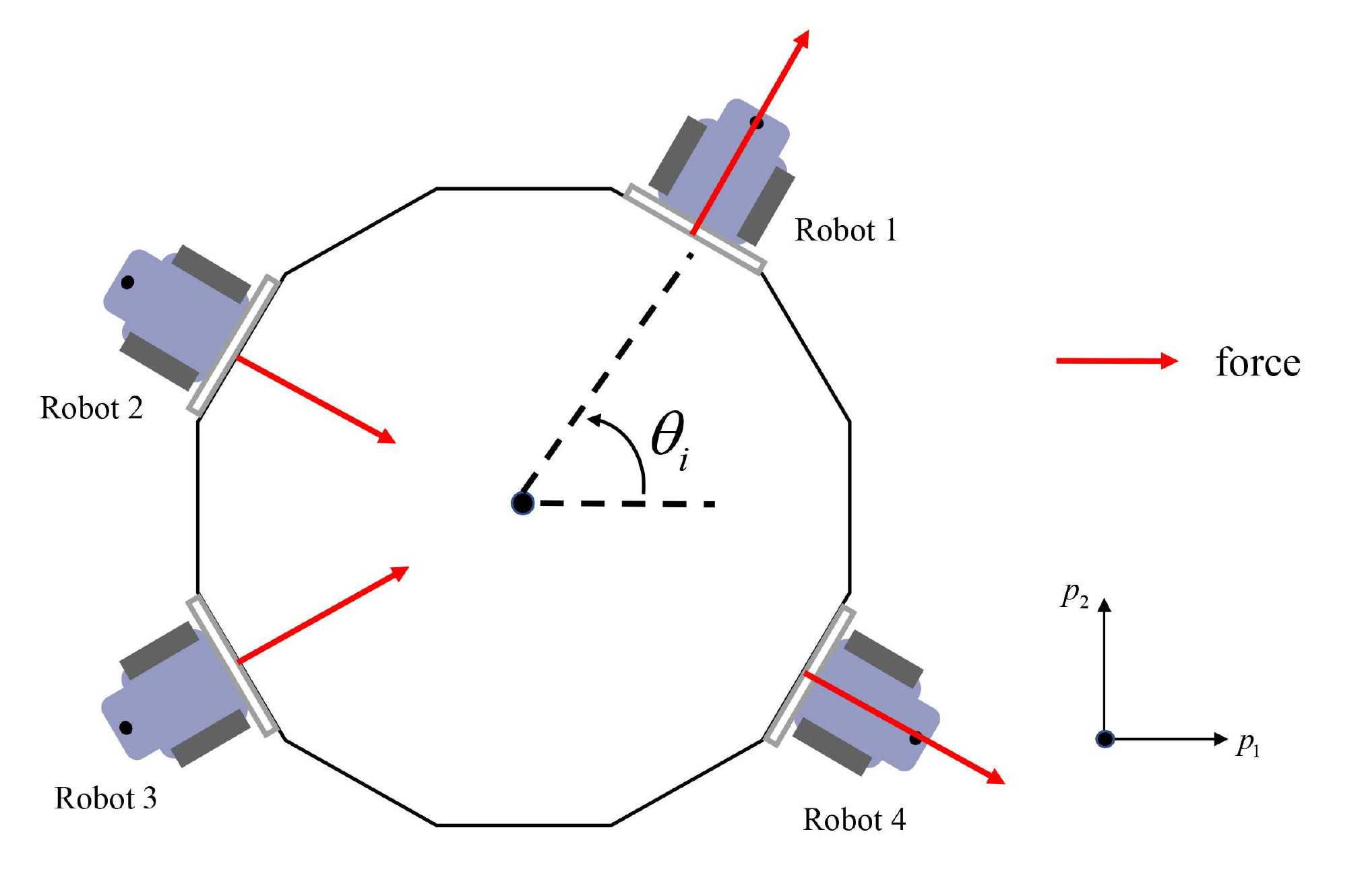}}}
\caption{Four robots carry an object.} \label{f:transportation}
\end{figure}

In this section, the cooperative transportation of an object by a network of mobile robots is considered. The transportation model is modified from \cite{kim2020decentralized} and described by (\ref{equ:model_s}), where $s_k \in \mathbb{R}^{4}$ denotes the planar position and velocity error of the object and $u_i \in \mathbb{R}$ denotes the pushing or pulling force exerted by robot $i$, $i=1$, $\ldots$, $4$, as illustrated in Fig.~\ref{f:transportation}.

Moreover, $s_k = [p_1 (k)- \bar p_1$, $p_2 (k)- \bar p_2$, $v_1 (k)$, $ v_2 (k)]^T $, where $p_1 (k)$, $p_2 (k)$, $v_1 (k)$ and $ v_2 (k)$ are the position and velocity of the object at step $k$; and $\bar p_1 $ and $\bar p_2$ are the desired position. The communication topology of the four robots is a directed circle, i.e., robot $1 \rightarrow$ robot $2 \rightarrow$  robot $3 \rightarrow$  robot $4 \rightarrow$  robot $1$. This setting shows that the proposed control strategy is also applied to directed communication topologies. Moreover, the model parameters are chosen as

\begin{align}
A =  &  \left [
  \begin{array}{c c c c }
  { 1  } & {0} & {T}  & {0} \\
  { 0  } & {1} & {0}  & {T} \\
  { 0  } & {0} & {1}  & {0} \\
  { 0  } & {0} & {0}  & {1}   \end{array}  \right ], 
 C =    \left [
  \begin{array}{c c}
  { C_1 }  \\
  { C_2 }   \\
  { C_3 }   \\
  { C_4 }   \end{array}  \right ]  =   \left [
  \begin{array}{c c c c }
  { 1  } & {0} & {0}  & {0} \\
  { 0  } & {1} & {0}  & {0} \\
  { 0  } & {0} & {0}  & {0} \\
  { 0  } & {0} & {0}  & {0}   \end{array}  \right ], \notag
\end{align}
and $B_i =  1/M \times [0 , \ 0, \ \cos \theta_i, \ \sin \theta_i ]^T$,  where $T=0.02 \ s$ is the sampling time, $M=5$ is the inertial of the object, and $\theta_i$ is the angle of the force. Besides,  $\theta_1 = \pi/4$, $\theta_2 = \pi/3$, $\theta_3 =  \pi/2$ and $\theta_4 = 3\pi/4$, where $\pi$ is  the ratio of the circumference of a circle to its diameter. The initial position and velocity of the object are set as $ [120$, $200$, $0$, $0]^T$.  The desired position is set as $[20$, $50]^T$. The parameters in Algorithm \ref{algorithm1} are chosen as  $\delta=0.001$, $M_2 = 20$, $\epsilon = 0.1$, $\pi_1 = 0.13$, $\pi_2 = 0.25$, $\pi_3  = 0.33$ and $\pi_4 = 0.42$.

\begin{figure}[t]
\center
\subfigure{{\includegraphics[scale=0.4]{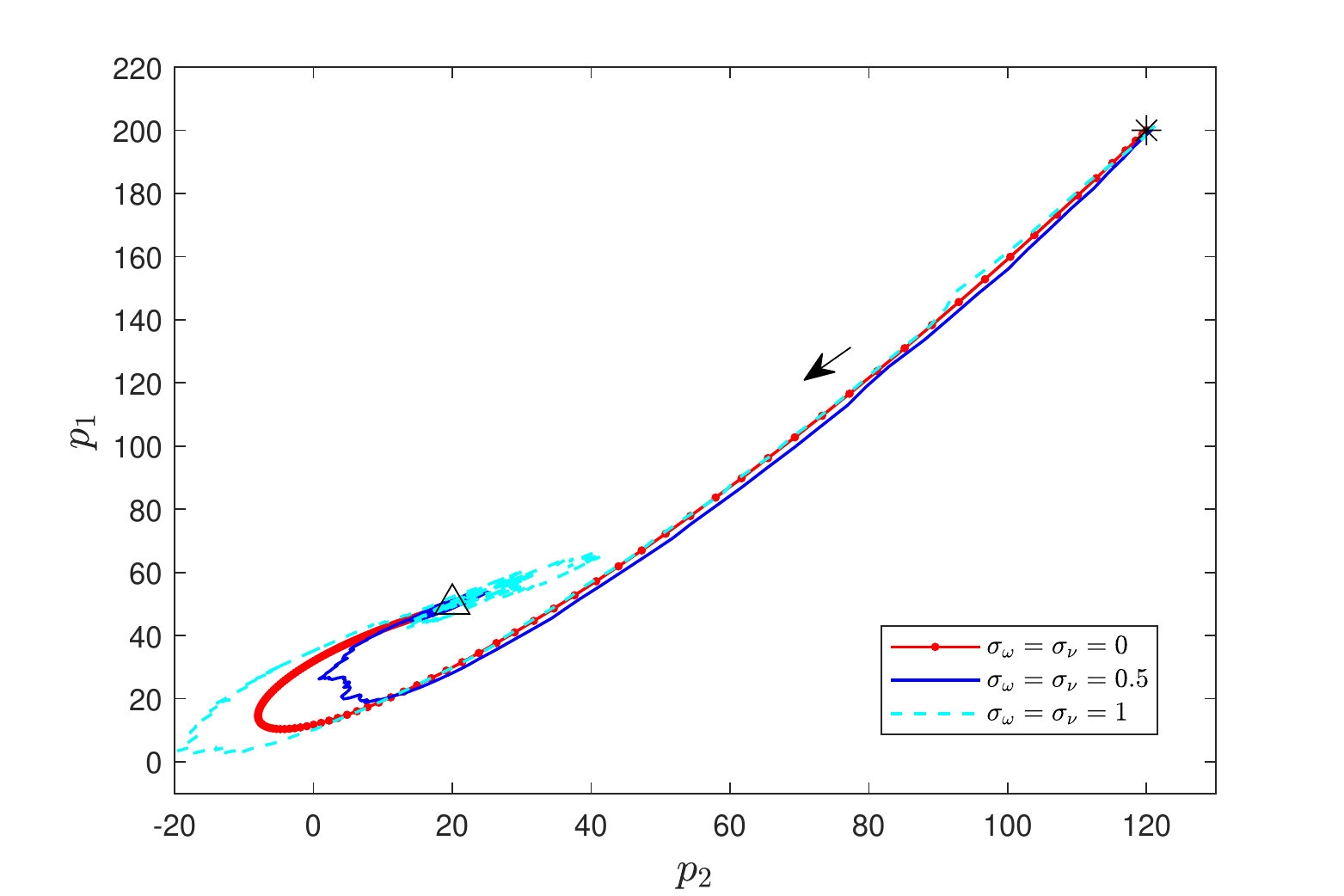}}}
\caption{{The trajectory of the object with different magnitudes of system noise, where '$*$' and '$\bigtriangleup$' denote the initial and terminal positions, respectively. }}\label{f:trajectory}
\end{figure}



\begin{figure}[t]
  \center
  \subfigure[The position errors of the object under the proposed strategy (\ref{equ:model_zi}) and (\ref{equ:model_ui}) without privacy preservation, $ M_1 = 10$.]{\includegraphics[scale=0.45]{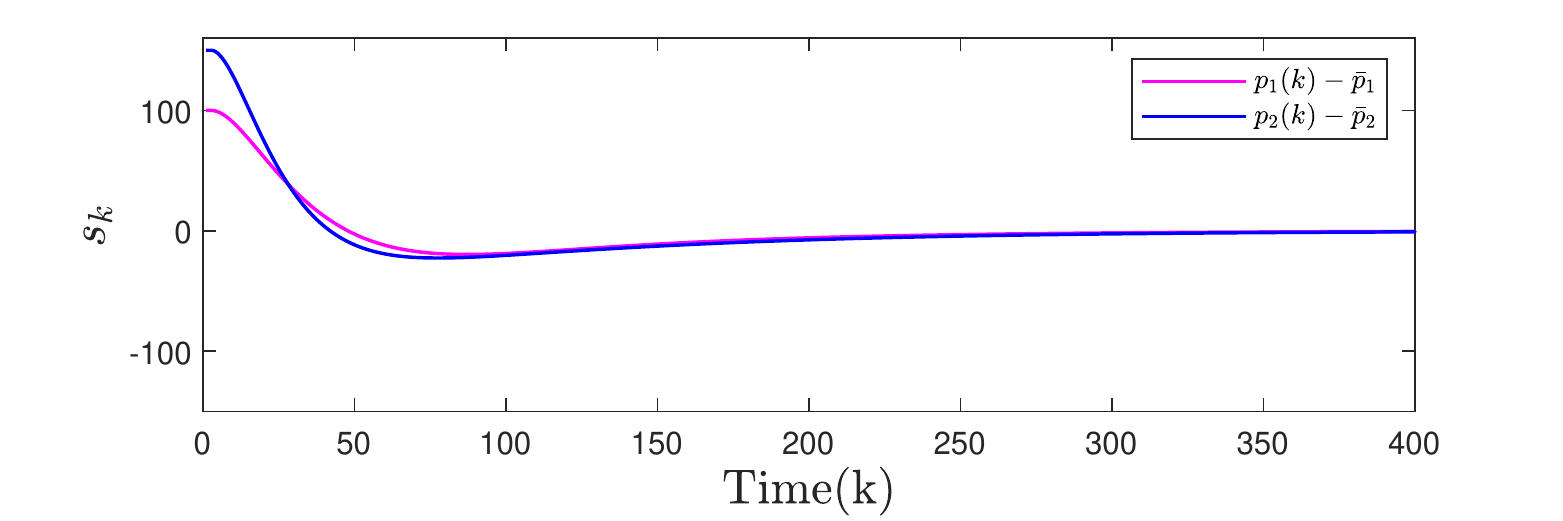}}
  \subfigure[The position errors of the object under Algorithm \ref{algorithm1} with privacy preservation, $ M_2 = 10$.]{\includegraphics[scale=0.45]{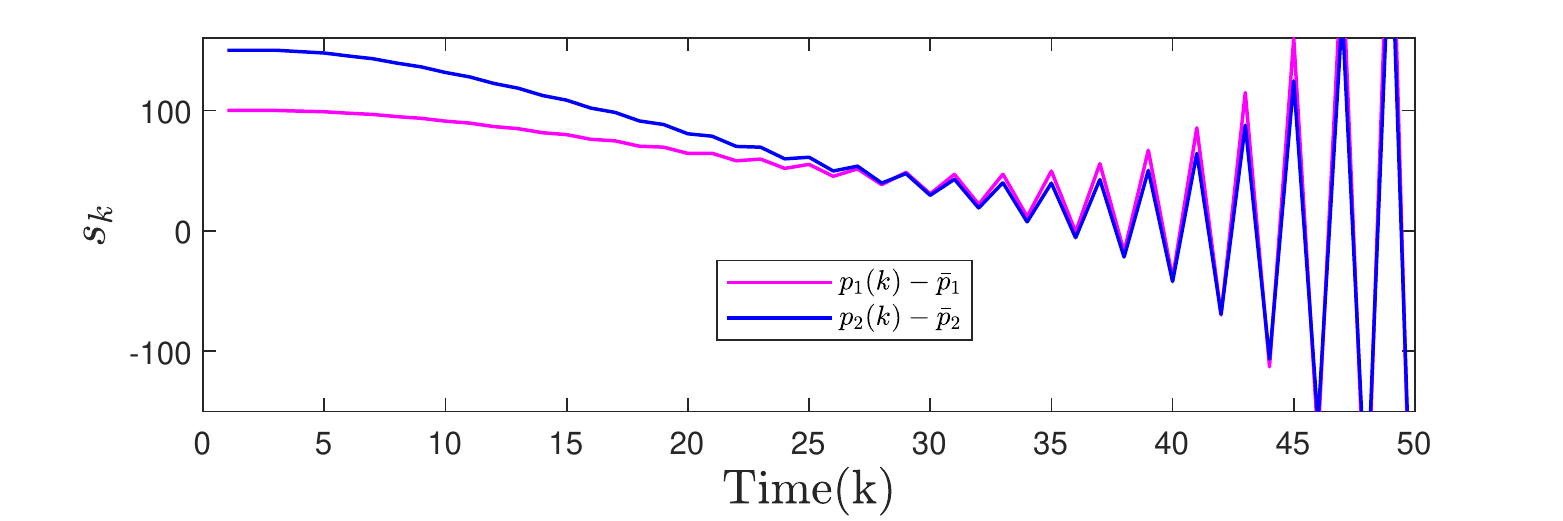}}
  \subfigure[The position errors of the object under Algorithm \ref{algorithm1} with privacy preservation, $ M_2 = 15$.]{\includegraphics[scale=0.45]{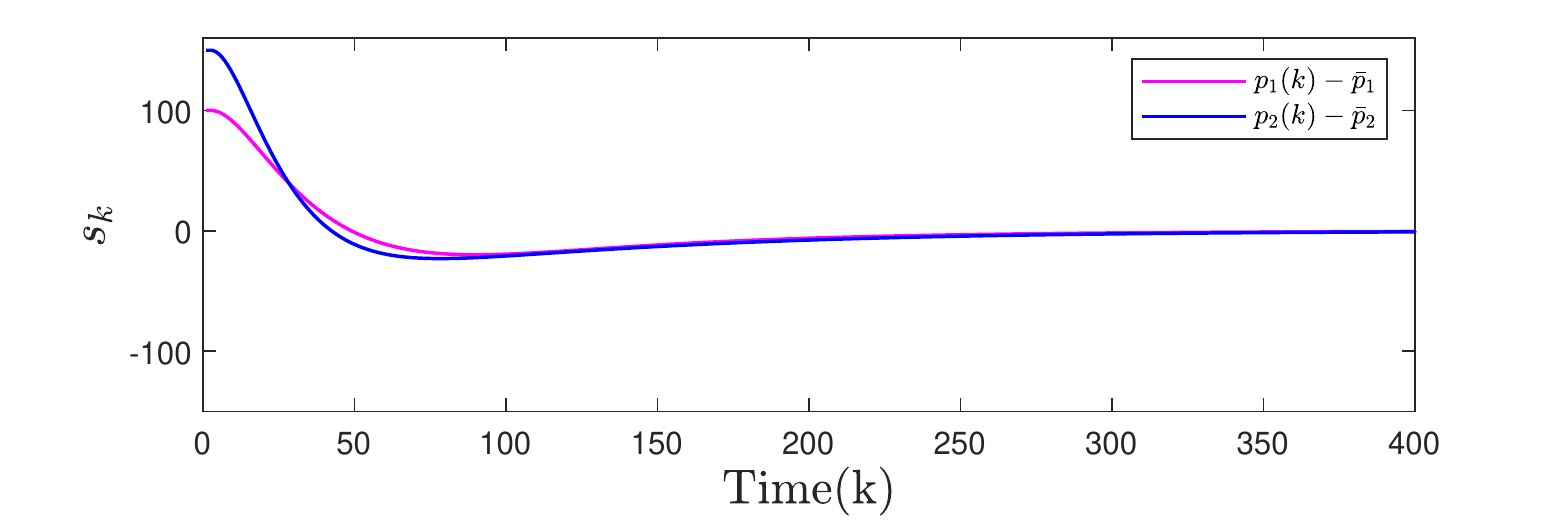}}
  \caption{The position errors under different algorithms and consensus steps.  } \label{f:comM}
  \end{figure}

  \begin{table}[ht]
    \centering 
    \caption{{The estimates of $\theta_1$ by Robot $2$ with different $\pi_1$.}}
    \label{tab:value}
    \centering
    \begin{tabular}{c|c|c|c|c|c|c}
    \hline
     $\pi_1$ & 0.1 & 0.2 & 0.3 & 0.4 & 0.5 & 0.6 \\
    \hline
    $\theta_1$ & 0.40 & 1.09 &  1.26 & 1.34 & 1.39 & 1.42 \\
    \hline
    \end{tabular}
    \end{table}

{The control performance of Algorithm \ref{algorithm1} is shown in Fig.~\ref{f:trajectory}, where the object is successfully transported from the initial position to the desired position. This outcome serves as a validation of the effectiveness and robustness of Algorithm \ref{algorithm1}. Next, we demonstrate the  privacy-preserving performance of Algorithm \ref{algorithm1}. We assume that Robot $2$ is a honest-but-curious adversary and attempts to infer the private angle $\theta_1$ of Robot $1$, but it cannot get access to the exact value of $\pi_1$. Since all pairs of $(\pi_1$, $\theta_1)$ in Table \ref{tab:value} can correspond to the same data $I_{12}$ received by Robot 2, which indicates that Robot $2$ cannot infer the unique private information of Robot $1$ with any guaranteed accuracy.} 
Moreover, the communication cost of the privacy-preserving mechanism is illustrated in Fig.~\ref{f:comM}. Note that the proposed algorithm (\ref{equ:model_zi}) and (\ref{equ:model_ui}) without privacy preservation can guarantee system stabilization when the fusion steps during each control interval are $10$, while Algorithm \ref{algorithm1} with privacy preservation requires 15 steps to achieve a similar performance. Hence, the incorporation of the privacy-preserving mechanism leads to a higher communication overhead, which fits with the theoretical result revealed in Theorem \ref{thm4}. In addition, the benefit of adding more channels/agents is revealed as follows. One more robot participates in the transportation task, denoted by robot $5$. The parameters for this robot are set as $\theta_5 = \pi $, $C_5 = [0,0,0,0]$ and $\pi_5 =0.55$. Other settings are the same as those in the case of four robots. Then, let $J_4$ and $J_5$ denote the normalized LQR index defined in (\ref{equ:lqrindex}) under Algorithm \ref{algorithm1}  with four robots and five robots, respectively. It can be computed that $J_4 = 9.5 \times 10^5$ and $ J_5 = 9.1 \times 10^5$, which indicates that adding more robots can improve the LQR performance under the proposed control framework. Hence, the effectiveness of Theorem \ref{thm4} is verified.
 
Thus, the effectiveness of the theoretical results in this paper has been illustrated.

\section{Conclusion}\label{sec5}
In this paper, a cooperative stabilization problem for multi-channel linear systems has been studied, where each channel is managed by an agent. A novel privacy-preserving state estimation and control algorithm has been designed, where agents can self-organize the algorithm with all private information preserved. In addition, the cost of introducing the privacy-preserving mechanism and the effect of adding more channels have been analyzed. The effectiveness of the proposed cooperative strategy has been illustrated by numerical examples. Future research focus will be on more complex cooperation tasks and security issues.

\appendices

\section{The proof of Theorem \ref{thm1}} \label{thm1proof}
It follows from (\ref{equ:model_zi}) and (\ref{equ:model_ui}) that
\begin{align}
 x_{k+1,0}^i = & \ ( A - N L^{i} C^i + N B^{i} K^i ) z_{k}^i  + N L^{i} C^i s_k. \notag
\end{align}
By denoting $x_{k+1,0} = [ (x_{k+1,0}^1)^T$, $\ldots$, $ (x_{k+1,0}^N)^T ]^T$, we have
\begin{align}
 x_{k+1,0}  = & \ \mathcal{A} z_{k}   + N \bar{\mathcal{L}}  \mathcal{C}  s_k, \notag
\end{align}
where $z_k = [ (z_{k}^1)^T$, $\ldots$, $(z_{k}^N)^T ]^T$, and $\mathcal{A}$, $\bar{\mathcal{L}}$ and $ \mathcal{C}$ are given in (\ref{equ:mathcalA}) and Assumption \ref{asm:detectability}, respectively. Then, according to (\ref{equ:model_zi}), we have
\begin{align}
 x_{k+1,l}  = & \ ( W^l \otimes I_n)  \mathcal{A} z_{k}   + N ( W^l \otimes I_n) \bar{\mathcal{L}}  \mathcal{C}  s_k , \notag
\end{align}
where $x_{k+1,l} = [ (x_{k+1,l}^1)^T$, $\ldots$, $ (x_{k+1,l}^N)^T ]^T$, and $W$ is defined in Section \ref{sec2.2}. Further, it follows from (\ref{equ:model_zi}) that
\begin{align} \label{equ:augmentz}
 z_{k+1}  = & \ ( W^{M_1} \otimes I_n)  \mathcal{A} z_{k}   + N ( W^{M_1} \otimes I_n) \bar{\mathcal{L}} \mathcal{C}  s_k.
\end{align}
To proceed, define $\eta_k = [s_k^T$, $z_k^T]^T$. According to (\ref{equ:model_s}) and (\ref{equ:augmentz}), the dynamics of $\eta_{k+1}$ can be derived as
\begin{align} \label{equ:eta1}
 \eta_{k+1} = &   F_1 \eta_k,
\end{align}
where
\begin{align} \label{equ:F1}
  F_1 = &  \left[{
  \begin{array}{*{20}{c}}
  {A} & {\mathcal{B}  \bar{\mathcal{K}} }   \\
  { N ( W^{M_1} \otimes I_n) \bar{\mathcal{L}}  \mathcal{C} } & {( W^{M_1} \otimes I_n)  \mathcal{A}}  \\
  \end{array}} \right],
\end{align}
with $\mathcal{B}$ and $ \bar{\mathcal{K}}$ being defined in Assumption \ref{asm:stabilizability} and  (\ref{equ:mathcalA}), respectively.

Now, it suffices to prove that $\eta_k$ is stable when the condition   (\ref{equ:M1}) holds. First, some features of $W$ are revealed. According to \cite{ren2005consensus}, when the communication topology is undirected and connected, there exists a unitary matrix $U = [\frac{1}{\sqrt{N}} \mathbb{1}_N, \  Y^T]^T \in \mathbb{R}^{N \times N} $ that $UU^T = I_N$ and $UWU^T = \text{diag} \{1$, $\Lambda \}$ with $\Lambda \in   \mathbb{R}^{(N-1) \times (N-1)}$ being a diagonal matrix. Moreover, $YW=\Lambda Y$, $Y \mathbb{1}_N =0_{N \times 1}$, $YY^T=I_{N-1}$, and $Y^T Y + \frac{1}{N} \mathbb{1}_N \mathbb{1}_N^T =I_{N}$. Then, introduce an auxiliary matrix as
\begin{align}
  Q_1  =  &  \left[{
  \begin{array}{*{20}{c}}
  {I_n} & {  }  \\
  { } & {(\frac{1}{\sqrt{N}}U) \otimes I_n}
  \end{array}} \right]. \notag
\end{align}
According to the definition of $U$, $Q_1^{-1}$ can be obtained as
\begin{align}
  Q_1^{-1}  =  &  \left[
  \begin{array}{*{20}{c}}
  {I_n} & {  }   \\
  { } & {( \sqrt{N} U^T ) \otimes I_n}
  \end{array}  \right]. \notag
\end{align}
By pre-multiplying and post-multiplying $F_1$ by $Q_1$ and $ Q_1^{-1}$, respectively, we have
\begin{align}
 &  Q_1 F_1 Q_1^{-1} = \notag \\
   & \bigg [
  \begin{array}{c c}
  {A} & {\sqrt{N} \mathcal{B}  \bar{\mathcal{K}} (U^T   \otimes I_n) }   \\
  {  \sqrt{N} ( U  W^{M_1} \otimes I_n) \bar{\mathcal{L}} \mathcal{C} } & { ( U W^{M_1} \otimes I_n)  \mathcal{A} ( U^T  \otimes I_n )}    \end{array}  \bigg ].   \notag
\end{align}
By substituting  $U = [\frac{1}{\sqrt{N}} \mathbb{1}_N, \  Y^T]^T  $ into the above equation and noting that $Y W^{M_1}  =\Lambda^{M_1} Y$, we have (\ref{equ:qfq2}).
\begin{figure*}[t]
\begin{align}  \label{equ:qfq2}
  &  Q_1 F_1 Q_1^{-1} =   \left [
  \begin{array}{c c c}
  {A} & {\sum_{i=1}^{N}  B^{i} K^{i}} & {\sqrt{N} \mathcal{B}  \bar{\mathcal{K}} (Y^T   \otimes I_n) }  \\
  {  \sum_{i=1}^{N}  L^{i} C^{i}} & {  A - \sum_{i=1}^{N}  L^{i} C^{i}  + \sum_{i=1}^{N}  B^{i} K^{i} } &
  { \frac{1}{\sqrt{N}} (\mathbb{1}_N^T \otimes I_n)  \mathcal{A} ( Y^T  \otimes I_n )}  \\
   {  \sqrt{N} ( \Lambda^{M_1} Y \otimes I_n) \bar{\mathcal{L}} \mathcal{C}} & {\frac{1}{\sqrt{N}} (  \Lambda^{M_1} Y   \otimes I_n )  \mathcal{A} (\mathbb{1}_N \otimes I_n)  } &  { ( \Lambda^{M_1} Y \otimes I_n)  \mathcal{A} ( Y^T  \otimes I_n )}   \end{array}  \right ]
\end{align}
\hrulefill
\end{figure*}
Define $R_1$ as
\begin{align}
 &  R_1 =  \left [
  \begin{array}{c c :c}
  {I_n} & { } & {  }  \\
  { - I_n} & { I_n } &   {  }  \\ \hdashline
   { } & {   } &  { I_{n(N-1)} }   \end{array}  \right ]. \notag
\end{align}
Pre-multiplying and post-multiplying $ Q_1 F_1 Q_1^{-1}$ in (\ref{equ:qfq2}) by $R_1$ and $ R_1^{-1}$, respectively, give
\begin{align}  \label{equ:qfq3}
  &  (R_1 Q_1) F_1 (R_1 Q_1)^{-1}
   =    \left [
  \begin{array}{c c}
  {F} & {  \Delta_{1} }  \\
  {   \Delta_{2} } & {  \Delta_{3}  }    \end{array}  \right ] \triangleq H_1,
\end{align}
where $F$ is given in (\ref{equ:F}) and
\begin{align}
   \Delta_{1}
   =  &  \frac{1}{\sqrt{N}}   \left [
  \begin{array}{c c}
  {   N (\mathbb{1}_N^T \otimes I_n) ( \bar{\mathcal{B}}  \bar{\mathcal{K}})   }  \\
  {     (\mathbb{1}_N^T \otimes I_n)( \mathcal{A} - N \bar{\mathcal{B}} \bar{ \mathcal{K}} )   }    \end{array}  \right ] (Y^T   \otimes I_n) , \notag \\
 \Delta_{2}
   =  & \frac{1}{\sqrt{N}} ( \Lambda^{M_1} Y \otimes I_n)  [
   { (\mathcal{A} + N \bar{\mathcal{L}} \bar{\mathcal{C}})(\mathbb{1}_N \otimes I_n) }   \ {  \mathcal{A}   }    (\mathbb{1}_N \otimes I_n)], \notag \\
  \Delta_{3}   =  &  ( \Lambda^{M_1} Y \otimes I_n)  \mathcal{A} ( Y^T  \otimes I_n ). \notag
\end{align}
Define a new variable $\xi_{k} = R_1 Q_1 \eta_k $. It follows from (\ref{equ:eta1}) that
\begin{align} \label{equ:xi1}
 \xi_{k+1} = &   H_1 \xi_k,
\end{align}
where $ H_1$ is given in (\ref{equ:qfq3}). Since $R_1 Q_1$ is nonsingular, it suffices to prove that $\xi_k$  is stable when the condition (\ref{equ:M1}) holds. To do so, introduce a Lyapunov function candidate as follows
\begin{align} \label{equ:Lyapunov1}
 V_{k} = & \xi_k^T  \left [
  \begin{array}{c c}
  {\delta P} & { }   \\
  { } & {  I_{n(N-1)} }    \end{array}  \right ] \xi_k,
\end{align}
where $P$ is given in (\ref{equ:lyapunove}) and $\delta$ is a small positive constant defined as
\begin{align} \label{equ:delta}
  \delta  = \frac{\theta}{2   \psi  (1 + \theta) \lambda_{\text{max}} (P)}
\end{align}
with $\theta$ and $\psi$ being given in (\ref{equ:lyapunove}) and (\ref{equ:M1}), respectively. Since $V_{k} \ge 0$, we only need to prove that $V_{k}$ is monotonically decreasing along the system (\ref{equ:xi1}). It follows from (\ref{equ:qfq3}) and (\ref{equ:xi1}) that
\begin{align}
    V_{k+1}
\leq &   \ \xi_{k}^T   \left (  (1+\theta) \left [
  \begin{array}{c c}
  { \delta F^T P F} & {  }  \\
  {     } & {     \Delta_{3}^T  \Delta_{3}  } \end{array}  \right ]   \right. \notag \\
    + & \left. (1+\theta^{-1})
   \left [
  \begin{array}{c c}
  {  \Delta_{2}^T \Delta_{2}  } & {   }  \\
  {  } & {\delta \Delta_{1}^T P \Delta_{1}  } \end{array}  \right ] \right )      \xi_{k} , \notag
\end{align}
where the inequality is obtained by using Young's inequality. By utilizing (\ref{equ:lyapunove}),  we have
\begin{align} \label{equ:star12}
   V_{k+1} -  V_{k} \leq     \xi_{k}^T    \left [
  \begin{array}{c c}
  { *_1   } & {  }  \\
  {     } & {  *_2  } \end{array}  \right ]     \xi_{k},
\end{align}
where
\begin{align}
*_1  &  =  - \delta Q + (1+\theta^{-1}) \Delta_{2}^T \Delta_{2},     \notag \\
*_2  &  = (1+\theta)  \Delta_{3}^T  \Delta_{3} +   (1+\theta^{-1}) \delta \Delta_{1}^T P \Delta_{1} - I_{n(N-1)}. \notag
\end{align}
Till now, it suffices to prove that $*_1 < 0$ and  $*_2 < 0$. Noting that $ \|Y\|_2  = \|Y^T\|_2  =  1$, $\|\mathbb{1}_N \|_2 = \|\mathbb{1}_N^T \|_2 = \sqrt{N}$,
it follows from   (\ref{equ:M1}) and (\ref{equ:qfq3}) that terms $\Delta_{1}^T P \Delta_{1}$, $\Delta_{2}^T \Delta_{2}$ and $\Delta_{3}^T  \Delta_{3}$ in $*_1$ and $*_2$ satisfy
\begin{align}
   \Delta_{1}^T P \Delta_{1} & \leq  \| P \|_2   (  \|  N \bar{\mathcal{B}}   \bar{\mathcal{K}} \|_2^2 +  \| \mathcal{A} -  N \bar{ \mathcal{B}}   \bar{\mathcal{K}} \|_2^2)  I_{n(N-1)}\notag \\
  & \leq \psi \lambda_{\text{max}} (P)  I_{n(N-1)},  \notag \\
 \Delta_{2}^T \Delta_{2}  & \leq  \lambda^{2 M_1} (  \|\mathcal{A} + N \bar{\mathcal{L}}  \bar{\mathcal{C}} \|_2^2 + \| \mathcal{A} \|_2^2)  I_{2n}
   \leq   \psi \lambda^{2 M_1}   I_{2n} , \notag \\
 \Delta_{3}^T  \Delta_{3} & \leq  \lambda^{2 M_1}  \| \mathcal{A} \|_2^2  I_{n(N-1)}
  \leq \psi  \lambda^{2 M_1}  I_{n(N-1)}, \notag
\end{align}
respectively, where $\lambda$ is the second largest eigenvalue of $W$. Further, according to (\ref{equ:M1}), we have
\begin{align} \label{equ:lambda2m}
\lambda^{2 M_1} <   \frac{\theta^2  }{2 (1 + \theta)^2 \psi^2  } \frac{ \lambda_{\text{min}}(Q)}{  \lambda_{\text{max}}(P) } .
\end{align}
Moreover, since $\psi \ge 1$, $\theta \in (0, \ 1)$ and $P \ge Q$, $\lambda^{2 M_1}$ in (\ref{equ:lambda2m}) can be relaxed by
\begin{align} \label{equ:lambda2mmm}
\lambda^{2 M_1} <  \min \bigg \{  \frac{1}{ 2 (1+\theta) \psi}, \  \frac{\theta^2 \lambda_{\text{min}}(Q)}{2 (1+\theta)^2 \psi^2 \lambda_{\text{max}}(P)} \bigg \},
\end{align}
Combined (\ref{equ:lambda2mmm}) with (\ref{equ:delta}), $*_1$ and $*_2$ in (\ref{equ:star12}) satisfy
\begin{align}
 *_1 & < -    \frac{\theta Q }{2   \psi  (1 + \theta) \lambda_{\text{max}} (P)}   +   \frac{\theta \lambda_{\text{min}}(Q) I_{2n}}{2 (1+\theta) \psi^2 \lambda_{\text{max}}(P)}  \leq 0, \notag
\end{align}
and
\begin{align}
 *_2 
  <  &  \bigg ( (1+\theta)  \psi   \frac{1}{ 2 (1+\theta) \psi}    - \frac{1}{2} \bigg ) I_{n(N-1)} \notag \\
  &  \qquad   +   \bigg ( (1+\theta^{-1})  \delta \psi \lambda_{\text{max}} (P)  - \frac{1}{2}  \bigg ) I_{n(N-1)}   
  =  0, \notag
\end{align}
respectively. Thus, the proof of Theorem \ref{thm1} is complete.

\section{The proof of Corollary \ref{cor1}} \label{cor1proof}

According to the proof of Theorem \ref{thm1}, it suffices to prove that the condition (\ref{equ:lambda2mmm}) holds if the condition (\ref{equ:M12}) holds.
To do so, first, we focus on $\|  N \bar{\mathcal{B}}   \bar{\mathcal{K}} \|_2$, $ \| \mathcal{A} -  N \bar{ \mathcal{B}}   \bar{\mathcal{K}} \|_2 $, $ \|\mathcal{A} + N \bar{\mathcal{L}}  \bar{\mathcal{C}} \|_2$ and $ \| \mathcal{A} \|_2 $ in (\ref{equ:M1}). When Assumption \ref{asm:boundness} holds, we have
\begin{align}
 \| N \bar{\mathcal{B}}   \bar{\mathcal{K}}  \|_2 & \leq N \kappa_B \kappa_K,  \notag \\
 \| \mathcal{A} -  N \bar{ \mathcal{B}}   \bar{\mathcal{K}}   \|_2  & \leq \kappa_A +   N  \kappa_L \kappa_C,  \notag \\
  \|\mathcal{A} + N \bar{\mathcal{L}}  \bar{\mathcal{C}} \|_2  & \leq \kappa_A +   N  \kappa_B \kappa_K,  \notag \\
  \|\mathcal{A}   \|_2  & \leq \kappa_A +   N  \kappa_B \kappa_K +  N  \kappa_L \kappa_C.  \notag
\end{align}
Then, by applying Young's inequality, $\psi$ in (\ref{equ:M1}) satisfies
\begin{align}
& \psi \leq \max \Big \{ 1, \  5 \kappa_A^2 +   5 N^2  \kappa_B^2 \kappa_K^2 +  3 N^2  \kappa_L^2 \kappa_C^2  \Big \} \triangleq   \psi_1. \notag
\end{align}
By replacing $\psi$ in (\ref{equ:lambda2mmm}) by $\bar \psi $, we have (\ref{equ:M12}), which indicates that (\ref{equ:M12}) is a sufficient condition of (\ref{equ:lambda2mmm}). Thus, the proof of Corollary \ref{cor1} is complete.

\section{The proof of Theorem \ref{thm2}} \label{thm2proof}

First, we denote the augmented form of $\alpha_{k,l}^i$, $\beta_{k,l}^i$, $z_{k,\alpha}^{i}$ and $z_{k,\beta}^{i}$ in (\ref{equ:model_zalphai}) by  
\begin{align}
 \alpha_{k,l} & = [(\alpha_{k,l}^1)^T, \ \ldots, \ (\alpha_{k,l}^N)^T]^T, \notag \\
  \beta_{k,l} & = [(\beta_{k,l}^1)^T, \ \ldots, \ (\beta_{k,l}^N)^T]^T, \notag \\
 z_{k,\alpha} & = [(z_{k,\alpha}^{1})^T, \ \ldots, \ (z_{k,\alpha}^{N})^T]^T, \notag \\
 z_{k,\beta} & = [(z_{k,\beta}^{1})^T, \ \ldots, \ (z_{k,\beta}^{N})^T]^T, \  l=0, 1, \ldots, M_2. \notag
\end{align}
Let $\gamma_{k,l} = [\alpha_{k,l}^T$, $\beta_{k,l}^T]^T$. It follows from (\ref{equ:model_zalphai}) that  
\begin{align} \label{equ:augmodel_gammai} 
 \left \{ \begin{array}{l} 
  \gamma_{k+1,l}  =  (\tilde{W} \otimes I_n )  \gamma_{k+1,l-1} ,  \\
\gamma_{k+1,0}   =   ( V \otimes I_n) \mathcal{A} z_{k}   + N (V \otimes I_n)  \bar{\mathcal{L}}  \mathcal{C}  s_k,   \\
 z_{k}  = [ I_{Nn}  \quad   I_{Nn}   ] \gamma_{k,M_2}, 
 \end{array} \right .
\end{align}
where $ V$ and $  \tilde{W} $ are given in (\ref{equ:M2}).  Similarly to the proof of Theorem \ref{thm1}, define an augmented state as $\tilde{\eta}_k = [s_k^T$, $\gamma_{k,M_2}^T]^T$. Combining (\ref{equ:augmodel_gammai}) with (\ref{equ:model_ui}), we have
\begin{align} \label{equ:eta2}
 \tilde{\eta}_{k+1} = &  F_2 \tilde{\eta}_k,
\end{align}
where 
\begin{align} 
  F_2 = &  \left[{
  \begin{array}{*{20}{c}}
  {A} & {\mathcal{B}  \bar{\mathcal{K}}[  I_{Nn},  \quad   I_{Nn}   ] }   \\
  { N ( \tilde{W}^{M_2} V \otimes I_n)  \bar{\mathcal{L}} \mathcal{C} } & {   (\tilde{W}^{M_2} V \otimes I_n)   \tilde{\mathcal{A}}  }  \\
  \end{array}} \right], \notag 
\end{align}
with $\tilde{\mathcal{A}} = \mathcal{A} [ I_{Nn}, \  I_{Nn}   ] $. To guarantee the convergence of $s_k$, it suffices to prove that $F_2$ is Schur stable. Noting that each row/column in $\tilde{W}$ is summing to 1, there exists a unitary matrix $\tilde{U} = [\frac{1}{\sqrt{2N}} \mathbb{1}_{2N}, \  \tilde{Y}^T]^T \in \mathbb{R}^{2N \times 2N} $ that $\tilde{U} \tilde{U}^T = I_{2N}$ and $\tilde{U} \tilde{W}  \tilde{U}^T = \text{diag} \{1$, $\tilde{\Lambda} \}$ with $\tilde{ \Lambda} \in   \mathbb{R}^{(2N-1) \times (2N-1)}$ being a diagonal matrix. In addition, $\| \tilde{ \Lambda} \|_2 < 1$ when $\forall \epsilon \in (0, \ \frac{2}{3})$, which will be proved in the proof of Theorem  \ref{thm4}. Moreover, $\tilde{Y} \tilde{W}=\tilde{\Lambda} \tilde{Y}$, $\tilde{Y} \mathbb{1}_{2N} =0_{2N \times 1}$, $\tilde{Y} \tilde{Y}^T=I_{2N-1}$ and $\tilde{Y}^T \tilde{Y} + \frac{1}{2N} \mathbb{1}_{2N} \mathbb{1}_{2N}^T =I_{2N}$. Now, introduce two auxiliary matrices
\begin{align}
  Q_2  =  &  \left[{
  \begin{array}{*{20}{c}}
  {I_n} & {  }  \\
  { } & { \sqrt{\frac{2}{N}} \tilde{U}  \otimes I_n}  \\
  \end{array}} \right],
    R_2 =  \left [
  \begin{array}{c c :c}
  {I_n} & { } & {  }  \\
  { - I_n} & { I_n } &   {  }  \\ \hdashline
   { } & {   } &  { I_{2Nn-n} }   \end{array}  \right ] \notag
\end{align}
respectively.
Further, noting that $ \tilde{U}  \tilde{W}^{M_2}  = \left[
  \begin{array}{*{20}{c}}
  {\frac{1}{\sqrt{2N}} \mathbb{1}_{2N}^T}    \\
  {\tilde{\Lambda}^{M_2} \tilde{Y}}  \\
  \end{array}  \right] $ and $V = \left [
  \begin{array}{c c}
  { \Pi  }   \\
  { (I_{N} -   \Pi)  } \end{array}  \right ]$, we have
\begin{align}  \label{equ:tildeqfq2}
  &  (R_2 Q_2) F_2 (R_2 Q_2)^{-1}
   =    \left [
  \begin{array}{c c}
  {F} & {  \tilde{\Delta}_{1} }  \\
  {   \tilde{\Delta}_{2} } & {  \tilde{\Delta}_{3}  }    \end{array}  \right ] \triangleq H_2,
\end{align}
where $F$ is given in (\ref{equ:F}) and
\begin{align}
   \tilde{\Delta}_{1}
   =  & \frac{1}{ \sqrt{2N}}   \left [
  \begin{array}{c c}
  {    (\mathbb{1}_N^T \otimes I_n) (N \bar{\mathcal{B}}  \bar{\mathcal{K}})   }  \\
  {     (\mathbb{1}_N^T \otimes I_n) (\mathcal{A} - N \bar{\mathcal{B}} \bar{ \mathcal{K}} )   }    \end{array}  \right ] [I_{Nn}, \ I_{Nn}] (\tilde{Y}^T   \otimes I_n) , \notag \\
  \tilde{\Delta}_{2}
   =  & \sqrt{\frac{2}{N}} ( \Lambda^{M_2} \tilde{Y} V \otimes I_n)  [
   { (\mathcal{A} + N \bar{\mathcal{L}} \bar{\mathcal{C}})  } , \quad {  \mathcal{A}   } ]    (\mathbb{1}_{2N} \otimes I_n), \notag \\
   \tilde{\Delta}_{3}   =  &  ( \Lambda^{M_2} \tilde{Y} V \otimes I_n)  \mathcal{A} [I_{Nn}, \ I_{Nn}]  ( \tilde{Y}^T  \otimes I_n ). \notag
\end{align}
By utilizing $\|(V \otimes I_n)  \mathcal{A} [I_{Nn}, \ I_{Nn}] \|_2 \leq \| \mathcal{A}  \|_2$ and  $\|[I_{Nn}, \ I_{Nn}] \|_2 \leq \sqrt{2}$, and following the similar derivation in Appendix \ref{thm1proof}, it can be proved that the closed-loop system (\ref{equ:model_s}) is stable if the condition (\ref{equ:M2}) holds. Thus, the proof of Theorem \ref{thm2} is complete.

 \section{The proof of Lemma \ref{thm3}} \label{thm3proof}
 First, denote $ \mathcal{B}_{M_3}  = [(\mathcal{B}^1)^T$, $\ldots$, $(\mathcal{B}^N )^T]^T$ and $ \mathcal{B}_v = [N B^{1} (B^{1})^T$, $\ldots$, $N B^{N} (B^{N})^T]^T$. It follows from (\ref{equ:model_Bi}) that
\begin{align}
   \mathcal{B}_{M_3}  = & [I_{nN},   \ I_{nN} ] ( \tilde{W} \otimes I_n )^{M_3} (V \otimes I_n )  \mathcal{B}_{v}, \notag
\end{align}
where $\tilde{W}$ and $ V$ are given in (\ref{equ:M2}). Further, it can be derived  that
\begin{align}
   & \mathcal{B}_{M_3} \! = \! [I_{nN},   \ I_{nN} ] ( \tilde{U}^T   \tilde{U}  \otimes I_n ) ( \tilde{W}^{M_3} \otimes I_n ) ( \tilde{U}^T   \tilde{U} V  \otimes I_n )   \mathcal{B}_{v}  \notag \\
    & = \! [I_{nN} ,  \ I_{nN} ] ( \tilde{U}^T   \otimes I_n ) (   \tilde{U} \tilde{W}^{M_3} \tilde{U}^T \otimes I_n ) (    \tilde{U} V \otimes I_n )    \mathcal{B}_{v} \notag \\
    & = \!  [I_{nN},  \ I_{nN} ]  \bigg ( \bigg [\frac{1}{\sqrt{2N}} \mathbb{1}_{2N}, \  \tilde{Y}^T \bigg ]   \otimes I_n \bigg )  \left [
  \begin{array}{c c}
  { I_n  } & {}  \\
  { } & {\tilde{\Lambda}^{M_3} \otimes I_n}  \end{array} \! \right ]    \notag \\
  & \qquad \times   \bigg ( \bigg [\frac{1}{\sqrt{2N}} \mathbb{1}_{2N}, \  \tilde{Y}^T \bigg ]^T   \otimes I_n \bigg )    ( V \otimes I_n )    \mathcal{B}_{v}  \notag \\
  &   = \! [I_{nN}, \ I_{nN} ]  \bigg ( \bigg [ \frac{ \mathbb{1}_{2N} \mathbb{1}_{2N}^T}{ 2N}  +  \tilde{Y}^T \tilde{\Lambda}^{M_3} \tilde{Y}  \bigg ]   \otimes I_n \bigg )  (   V \otimes I_n )    \mathcal{B}_{v},   \notag
\end{align}
where $\tilde{U} $ and $\tilde{\Lambda}$ are given below (\ref{equ:eta2}). Then, noting that
\begin{align}
& \frac{1}{ 2N}  [I_{nN},   \ I_{nN} ]   (    \mathbb{1}_{2N} \mathbb{1}_{2N}^T      \otimes I_n   )     (    V \otimes I_n )  = \frac{1}{N}    (\mathbb{1}_{N} \mathbb{1}_{N}^T   ) \otimes I_n , \notag
\end{align}
and $ \lim_{M_3 \rightarrow \infty} \tilde{\Lambda}^{M_3} = 0_{2N-1} $, we have
\begin{align}
   & \lim_{M_3 \rightarrow \infty}  \mathcal{B}_{M_3}   = \frac{1}{N}  [  (\mathbb{1}_{N} \mathbb{1}_{N}^T   ) \otimes I_n  ]  \mathcal{B}_{v}  =  \mathbb{1}_{N} \otimes  \mathcal{B} \mathcal{B}^T , \notag
\end{align}
Moreover, $\mathcal{B}^{i}$ converges to $\mathcal{B} \mathcal{B}^T$ exponentially as $M_3$ increases. Hence, with Lemma \ref{lemma1},  $(A, \ \mathcal{B}^i)$ is stabilizable and the Riccati equation below (\ref{equ:kd}) is feasible, rewritten as
\begin{align}
 P_K  =  & I_n -  A^T   P_K   \mathcal{B} \mathcal{B}^T (  \mathcal{B} \mathcal{B}^T P_K    \mathcal{B} \mathcal{B}^T + I_{n})^{-1}   \mathcal{B} \mathcal{B}^T P_K A \notag \\
  &    + A^T   P_K   A. \notag
 \end{align}
This ensures that $A -   \mathcal{B} \mathcal{B}^T  (I_{n} + \mathcal{B} \mathcal{B}^T P_K  \mathcal{B} \mathcal{B}^T)^{-1} \mathcal{B} \mathcal{B}^T  P_K A$ is Schur stable.
Substituting (\ref{equ:kd}) into $A + \sum_{i=1}^{N} B^{i} K^{i} $ yields
 \begin{align}
 & A + \sum_{i=1}^{N} B^{i} K^{i}  
  =  A -    \mathcal{B} \mathcal{B}^T  (I_{n} + \mathcal{B} \mathcal{B}^T P_K  \mathcal{B} \mathcal{B}^T)^{-1} \mathcal{B} \mathcal{B}^T  P_K A,  \notag
 \end{align}
which indicates that  $A + \sum_{i=1}^{N} B^{i} K^{i}$ is Schur stable. Similarly,  it can be proved that $\mathcal{C}^{i}$ converges to $ \mathcal{C}^T \mathcal{C}$ exponentially as $M_4$ increases and $A - \sum_{i=1}^{N}  L^{i} C^{i} $ is Schur stable.

\section{The proof of Theorem \ref{privacythm}} \label{privacythmproof}
{  First, we prove that adversary $\zeta$ cannot infer $B^i$ with any guaranteed accuracy if $\mathcal{N}_i \not \subset \mathcal{N}_{\zeta}$. According to Algorithm \ref{algorithm1}, it suffices to prove that adversary $\zeta$ cannot infer $B^{i}$ based on the data $I_{i \zeta}$ and the interaction laws (\ref{equ:model_zalphai}) and (\ref{equ:model_Bi}). On the one hand, we prove that adversary $\zeta$ cannot infer $B^{i}$ from (\ref{equ:model_Bi}). When $\mathcal{N}_i \not \subset \mathcal{N}_{\zeta}$, there exists at least one neighbor of agent $i$ that is not the neighbor of adversary $\zeta$. Without loss of generality, let one of such neighbors be denoted by agent $p$. Then, according to the model of curious adversaries, adversary $\zeta$ has no access to the information of agent $p$. Hence, adversary $\zeta$ does not know $\{\bar{\mathcal{B}}^{p}_{h}\}_{h=0}^{M_3-1}$. Then, it follows from (\ref{equ:model_Bi}) that adversary $\zeta$ has $2M_3$ constraint equations 
  \begin{align} 
    \left\{ \begin{array}{l} 
   \bar{\mathcal{B}}^{i}_{h} = \sum_{j \in \mathcal{N}_i} w_{ij} \bar{\mathcal{B}}^{j}_{h-1} - \epsilon \pi_i (\bar{\mathcal{B}}^{j}_{h-1} - \hat{\mathcal{B}}^{j}_{h-1}),  \\
   \hat{\mathcal{B}}^{i}_{h} = (1 - \epsilon  \pi_i ) \hat{\mathcal{B}}^{j}_{h-1} + \epsilon \pi_i \bar{\mathcal{B}}^{j}_{h-1}, \ h = 1, \dots, M_3.    \notag
  \end{array} \right. 
  \end{align} 
  However, even with the information $I_{i \zeta}$, there are at least $2M_3 + 1$ unknown variables in the above equations for adversary $\zeta$, i.e., $\{\bar{\mathcal{B}}^{p}_{h}, \ \hat{\mathcal{B}}^{i}_{h} \}_{h=0}^{M_3-1}$ and $\pi_i$. In this sense, it is impossible for adversary $\zeta$ to obtain the exact values of the unknown variables by solving above constraints. In other words, there exist infinite solutions of $\{\bar{\mathcal{B}}^{p}_{h}, \ \hat{\mathcal{B}}^{i}_{h}, \ \pi_i \}_{h=0}^{M_3-1}$ that correspond to the same data $I_{i \zeta}$. This further indicates that there exist infinite $B^{i}$ corresponding to the same data $I_{i \zeta}$ based on (\ref{equ:model_Bi}). Hence, adversary $\zeta$ cannot learn $B^{i}$ with $I_{i \zeta}$ and (\ref{equ:model_Bi}). On the other hand, it can be proved that adversary $\zeta$ cannot learn $B^{i}$ with $I_{i \zeta}$ and (\ref{equ:model_zalphai}) neither. 
  
  By adopting the similar inductive reasoning, we can prove that adversary $\zeta$ cannot infer other private information in $\mathcal{M}^i$. Thus, the proof of Theorem \ref{privacythm} is complete. }

  \section{The proof of Corollary \ref{corollarynoise}} \label{corollarynoiseproof}
  { 
    First, let $\nu_k =[(\nu_k^1)^T$, $\ldots$, $(\nu_k^N)^T]^T$. It follows from (\ref{equ:model_noise}) and Algorithm \ref{algorithm1} that the dynamics of $\breve{\eta}_k \triangleq [s_k^T$, $\gamma_{k,M_2}^T]^T$ defined above (\ref{equ:eta2}) can be derived as
    \begin{align} 
     \breve{\eta}_{k+1} = &  F_2 \breve{\eta}_k + F_\omega \breve{\omega}_k, \ k \in \mathbb{N},  
    \end{align}
    where $\breve{\omega} =[\omega_k^T$, $\nu_k^T]^T $ and other variables are defined the same as those in (\ref{equ:eta2}). Define a Lyapunov function candidate as $V_{k} = \mathbb{E} \{ \breve{\eta}_k^T \breve{P} \breve{\eta}_k \} $. According to (\ref{equ:lyapunovbreve}) and the above equation,  $V_{k}$ satisfies
    \begin{align}  
      V_{k} & =   \mathbb{E} \{  (F_2 \breve{\eta}_{k-1} + F_\omega \breve{\omega}_{k-1})^T \breve{P} (F_2 \breve{\eta}_{k-1} + F_\omega \breve{\omega}_{k-1}) \} \notag \\
      & \leq  \mathbb{E} \{ \breve{\eta}_{k-1}^T F_2^T \breve{P} F_2 \breve{\eta}_{k-1} \}  +   \lambda_{\text{max}} (F_\omega^T \breve{P} F_\omega )( n \sigma_{\omega}^2 + m \sigma_{\nu}^2 ) \notag \\
      & \leq \breve{\theta} V_{k-1} +    \lambda_{\text{max}} ( F_\omega^T \breve{P} F_\omega )( n \sigma_{\omega}^2 + m \sigma_{\nu}^2 ) . \notag  
     \end{align}
    By using mathematical induction, we have 
    \begin{align}  
      V_{k} \leq &  \breve{\theta}^k V_{0} +   \lambda_{\text{max}} ( F_\omega^T \breve{P} F_\omega )( n \sigma_{\omega}^2 + m \sigma_{\nu}^2 )   \sum_{h=0}^{k-1}\breve{\theta}^h . \notag  
     \end{align}
    Noting that $\breve{\theta} \in(0$, $1)$, it is derived that  
    \begin{align}  
      \lim_{k \rightarrow \infty}  V_{k} \leq &  \frac{ \lambda_{\text{max}} ( F_\omega^T \breve{P} F_\omega )( n \sigma_{\omega}^2 + m \sigma_{\nu}^2 )}{1 - \breve{\theta} },  \notag  
     \end{align}
    which indicates (\ref{equ:sbound}).  Thus, the proof of corollary \ref{corollarynoise} is complete. }

\section{The proof of Theorem \ref{thm4}} \label{thm4proof}

To show  $\bar M_1 < \bar  M_2$, it suffices to prove that $\lambda < \tilde{\lambda}$ and $\psi \leq \tilde{\psi}$. First, $\lambda$ and $\tilde{\lambda}$ is compared. Introduce an auxiliary stochastic matrix $\tilde{W}_1$ as 
\begin{align} 
     \tilde{W}_1 = \! & \! \left [
    \begin{array}{c c}
    {W  - \epsilon I_N } & {\epsilon I_N  }  \\
    {  \epsilon I_N  } & {  (1 - \epsilon ) I_N } \end{array}  \right  ], \notag 
  \end{align}
and let $\tilde{\lambda}_1$ denote the second largest absolute value of eigenvalues of $\tilde{W}_1$. Noting that $\tilde{W}_1 < \tilde{W}$ since $\Pi < I_N $, we have $\tilde{\lambda}_1 < \tilde{\lambda}$. Hence, it suffices to prove $\lambda \leq \tilde{\lambda}_1$ to guarantee $\lambda < \tilde{\lambda}$. Let the eigenvalues of $W$ be denoted by $\{ 1$, $\lambda_1 (= \lambda)$, $\lambda_2$, $\ldots$, $\lambda_{N-1} \}$, which are ranked in the descending order. Without loss of generality, $\Lambda$ defined in Appendix \ref{thm1proof} is represented as $\Lambda = \text{diag} \{\lambda_1$, $\ldots$, $\lambda_{N-1} \}$. Accordingly, the eigenvalues of $\tilde{W}_1$ can be directly derived as
\begin{align} \label{equ:tildeW1}
\lambda(\tilde{W}_1) = \frac{1+\lambda(W) \pm \sqrt{(1 - \lambda(W))^2 + 4 \epsilon^2 } }{2} - \epsilon, 
\end{align}
where $\lambda(W) \in \{ 1$, $\lambda_1$, $\lambda_2$, $\ldots$, $\lambda_{N-1} \}$. When $\lambda(W) \in [0,\ 1]$, we can prove that $\lambda(\tilde{W}_1) \in (-1,\ 1] $ and $\tilde{W}_1$ only has a simple eigenvalue $1$, $\forall \epsilon \in (0, \ \frac{2}{3})$. Next, denote the eigenvalues of $\tilde{W}_1$ by $\{ 1$, $\tilde{\lambda}_1 $, $\tilde{\lambda}_2$, $\ldots$, $\tilde{\lambda}_{2N-1} \}$, which rank in a descending order. Then, we can directly derive that  
\begin{align}
 \tilde{\lambda}_1 = \frac{1+\lambda  + \sqrt{(1 - \lambda )^2 + 4 \epsilon^2 } }{2} -  \epsilon > \lambda, \notag
\end{align}
for all $\epsilon \in (0, \ \frac{2}{3})$.
In the following, $\psi \leq \tilde{\psi}$ is ensured. It follows from (\ref{equ:M2}) that
\begin{align}
\| V \|_2^2 = \! \| \Pi \|_2^2 +   \|  I_{N} -   \Pi \|_2^2
  =   \!  \max_{i} \{ \pi_i^2 + (1-\pi_i)^2 \}    \ge \!  \frac{1}{2}.  \notag
\end{align}
Then, according to (\ref{equ:M1}) and (\ref{equ:M2}), we have $\psi \leq \tilde{\psi}$. Thus, the proof of Theorem \ref{thm4} is complete.

\section{The proof of Theorem \ref{thm5}} \label{thm5proof}
First, introduce three  matrix functions as
\begin{align}
 f_0 (X) = & (X^{-1} + \mathcal{B}_0 \mathcal{B}_0  )^{-1}, \notag \\
 f_1 (X) = & (X^{-1} + \mathcal{B}_1 \mathcal{B}_1  )^{-1}, \notag \\
 h (X) = & I_n + A^T X A.   \notag
 \end{align}
where $ \mathcal{B}_0$ and  $\mathcal{B}_1$ are given in Proposition \ref{pro1} and below (\ref{equ:model_snew}), respectively. Then, define four series of matrices as
\begin{align}
 P_{0,l}  =   h (  \bar P_{0,l-1}), \ \bar P_{0,l}  = & f_0 (  P_{0,l}),   \notag \\
 P_{1,l}  =   h (  \bar P_{1,l-1}), \ \bar P_{1,l}  = &  f_1 (  P_{1,l}) , \ l \in \mathbb{N}^{+},   \notag
 \end{align}
where  $ \bar P_{0,0} = \bar P_{1,0} = 0 $. It can be found that
 \begin{align}
 P_{0} = \lim_{l \rightarrow \infty} P_{0,l},  \  P_{1} = \lim_{l \rightarrow \infty} P_{1,l}.    \notag
 \end{align}
where $ P_{0}$ and $ P_{1}$ are given in Section \ref{sec3.5}. In the following,  $P_0 \ge P_1$ will be ensured by using mathematical induction. At step $l=1$, $P_{0,1}  =   h (  \bar P_{0,0}) = I_n \ge P_{1,1}  =   h (  \bar P_{1,0})  = I_n   $. If $P_{0,l}  \ge P_{1,l}  $, we have
 \begin{align}
\bar P_{0,l}^{-1} &  =    P_{0,l}^{-1} + \mathcal{B}_0 \mathcal{B}_0
 \leq P_{1,l}^{-1} + \mathcal{B}_1 \mathcal{B}_1  = \bar{P}_{1,l}^{-1},  \notag
 \end{align}
since $\mathcal{B}_0 \mathcal{B}_0 \leq \mathcal{B}_1 \mathcal{B}_1$. Further, we have $\bar P_{0,l} \ge \bar P_{1,l} $ and $P_{0,l} =  h (  \bar P_{0,l})  \ge  h (  \bar P_{1,l}) = P_{1,l}  $. By mathematical induction, we have $P_{0,l}   \ge   P_{1,l}  $, $\forall l \in \mathbb{N}^{+}$, which indicates that  $P_0 \ge P_1$. Hence, $J_0 \ge J_1$. Thus, the proof of Theorem \ref{thm5} is complete.

\bibliographystyle{IEEEtran}
\bibliography{ref}

\end{document}